\documentclass[aps,prb,twocolumn,unsortedaddress,superscriptaddress,showpacs]{revtex4-2}

\usepackage{multirow}
\usepackage{color}
\usepackage{graphicx}
\usepackage{amsmath,amsfonts,amsthm} 
\usepackage{physics,siunitx} 
\usepackage{booktabs,array}
\usepackage[T1]{fontenc} 
\usepackage{placeins}

\graphicspath{{./images/}}

\newcolumntype{C}[1]{>{\centering\arraybackslash}p{#1}}

\begin{document}

\title{Activating magnetoelectric optical properties\\  by twisting antiferromagnetic bilayers }

\author{Kunihiro Yananose}
\affiliation{Center for Theoretical Physics, Department of Physics and Astronomy, Seoul National University, Seoul 08826, Republic of Korea}
\author{Paolo G. Radaelli}
\affiliation{Clarendon Laboratory, Department of Physics, University of Oxford, Parks Road, Oxford OX1 3PU, United Kingdom}
\author{Mario Cuoco}
\affiliation{Consiglio Nazionale delle Ricerche, Institute for Superconducting and Innovative Materials and Devices (CNR-SPIN), c/o Universit\`a degli Studi di Salerno, I-84084 Fisciano (Salerno), Italy}
\author{Jaejun Yu}
 \email{jyu@snu.ac.kr}
\affiliation{Center for Theoretical Physics, Department of Physics and Astronomy, Seoul National University, Seoul 08826, Republic of Korea}
\author{Alessandro Stroppa}
 \email{alessandro.stroppa@spin.cnr.it}
\affiliation{Consiglio Nazionale delle Ricerche, Institute for Superconducting and Innovative Materials and Devices (CNR-SPIN), c/o Department of Physical and Chemical Sciences, Universit\`a degli Studi dell'Aquila, Via Vetoio I-67100 Coppito, L'Aquila, Italy}

\begin{abstract}
Twisting in bilayers introduces structural chirality with two enantiomers, \textit{i.e.}, left- and right-hand bilayers, depending on the oriented twist angle.
The interplay between this global chirality and additional degrees of freedom, such as magnetic ordering and the local octahedral chirality arising from the geometry of the bonds, can yield striking phenomena.
In this work, we focus on collinear antiferromagnetic CrI$_3$ twisted homo-bilayers, which are characterized by a staggered octahedral chirality in each monolayer. Using symmetry analysis, density functional theory and tight-binding model calculations we show that layers twisting can lower the structural and magnetic point-group symmetries,  thus activating pyroelectricity and the magneto-optical Kerr effect, which would otherwise be absent in untwisted antiferromagnetic homo-bilayers. Interestingly, both electric polarization and Kerr angle are controllable by the twist angle and their sign is reversed when switching from left- to right-twisted bilayers.  We further unveil the occurrence of unconventional vortices with spin textures that alternate opposite chiralities in momentum space.
These findings demonstrate that the interplay between twisting and octahedral chirality in magnetic bilayers and related van der Waals heterostructures represents an extraordinary resource for tailoring their physical properties for spintronic and optoelectronic applications.
\end{abstract}

\date{\today}
\maketitle

\section{Introduction}

Monolayer CrI$_3$ is one of the first members of a newly-discovered class of two-dimensional (2D) ferromagnetic (FM) materials~\cite{huang_layer-dependent_2017}, which generated enormous interest, from both theoretical and experimental sides
\cite{liu_exfoliating_2016, sivadas_gate-controllable_2016, gong_discovery_2017, lado_origin_2017, samarth_magnetism_2017,  xu_interplay_2018, sivadas_stacking-dependent_2018, webster_strain-tunable_2018, huang_electrical_2018, fang_large_2018, burch_magnetism_2018, song_giant_2018, tong_skyrmions_2018, wu_physical_2019, kumar_gudelli_magnetism_2019, jang_microscopic_2019, molina-sanchez_magneto-optical_2020, yang_VI3_2020, yang_magneto-optical_2020, soriano_magnetic_2020, wang_stacking_2020, li_moire_2020, ghader_magnon_2020, hejazi_noncollinear_2020, sarkar_magnetic_2021, egorov_antiferromagnetism-induced_2021, jiang_recent_2021, song_direct_2021, lei_magnetoelectric_2021, akram_skyrmions_2021, ghader_whirling_2022, shen_exotic_2021, gibertini_magnetism_2021, kong_switching_2021, yu_interlayer_2021, xu_coexisting_2022, cheng_electrically_2022, fumega_moire-driven_2022, ghosh_chirality_2022}. 
In CrI$_3$, spin anisotropy enables ferromagnetic order to remain even in the monolayer limit by suppressing the fluctuation which is stated by the Mermin-Wagner theorem~\cite{huang_layer-dependent_2017, mermin_absence_1966}. In this respect, 2D magnetic materials allow precise investigations of truly 2D magnetism~\cite{samarth_magnetism_2017, burch_magnetism_2018, lado_origin_2017, xu_interplay_2018, jiang_recent_2021} and they show a remarkable tunability of their physical properties under external parameters such as strain~\cite{webster_strain-tunable_2018}, voltage gating~\cite{sivadas_gate-controllable_2016, lei_magnetoelectric_2021} or by inclusion in more complex heterostructures~\cite{yang_magneto-optical_2020}. 2D magnetic materials are also becoming an interesting platform to test concepts for future device applications in spintronics~\cite{song_giant_2018} and optoelectronics~\cite{sivadas_gate-controllable_2016}.

Twisting, that is, stacking two monolayer sheets with a rotational misalignment, represents an additional degree of freedom for tailoring the physical properties of 2D materials~\cite{cao_unconventional_2018, andrei_graphene_2020, hennighausen_twistronics_2021, lucignano_crucial_2019}. It is natural to expect that introducing twisting in magnetic bilayers would further increase their tunability, and may even produce new physical properties altogether. Working along this direction, several studies have already appeared in the literature focussing on moir\'e magnon bands~\cite{li_moire_2020, ghader_magnon_2020} or moir\'e-scale magnetic orderings~\cite{tong_skyrmions_2018, hejazi_noncollinear_2020, akram_skyrmions_2021, wang_stacking_2020, ghader_whirling_2022, song_direct_2021, xu_coexisting_2022, cheng_electrically_2022, fumega_moire-driven_2022}, while other studies have considered the effects of the electric~\cite{shen_exotic_2021, cheng_electrically_2022} or magnetic~\cite{soriano_spinorbit_2021} perturbations.

In this work, we focus on twisted magnetic bilayers in order to investigate whether twisting can lower symmetry in such a way as to activate new physical properties that are absent in the untwisted case. In particular, we study the possible occurrence of two properties that are key to many functional concepts: the magneto-optical Kerr effect (MOKE) and the electric polarization. MOKE refers to the rotation of the polarization axis of the light reflected from a magnetic material, and is a highly surface-sensitive technique, making it particularly suitable to investigate magnetic properties down to the monolayer limit~\cite{huang_layer-dependent_2017, huang_electrical_2018}. While all ferromagnets display MOKE signals, very recently the possibility that perfectly compensated magnets, \textit{e.g.}, antiferromagnets, may show Kerr rotation has been investigated~\cite{feng_large_2015, sivadas_gate-controllable_2016, higo_large_2018, zhou_spin-order_2019, yang_magneto-optical_2020, zhou_crystal_2021}. It has been known that the untwisted antiferromagnetic (AFM) CrI$_3$ bilayer is not MOKE active~\cite{huang_layer-dependent_2017, yang_magneto-optical_2020}.

\begin{figure}[t!]
\centering
\includegraphics[width=0.5\textwidth]{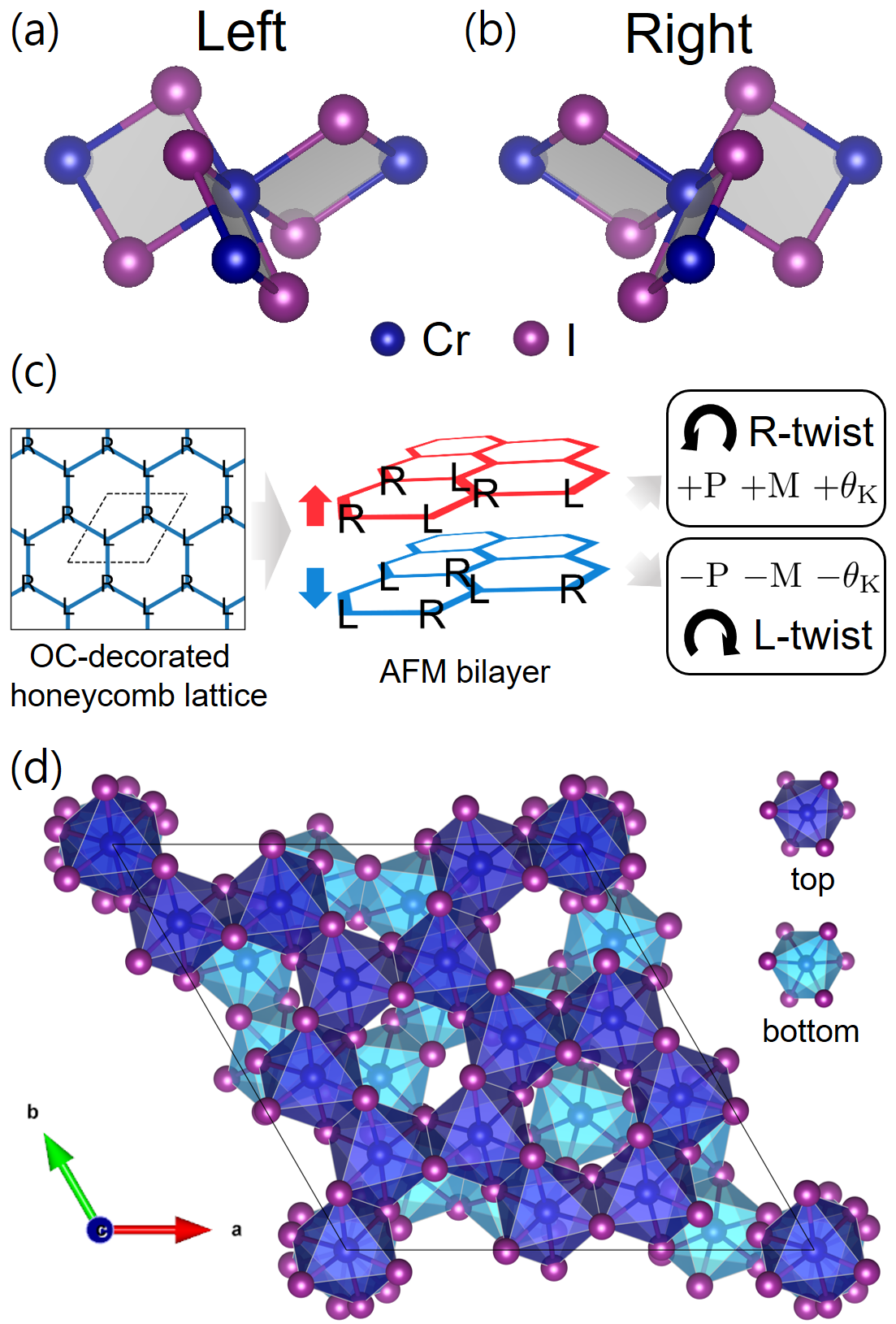}
\caption{
An enantiomeric octahedral complexes, (a) Left (L) and (b) Right (R). (c) A conceptual guide for this work. Two honeycomb lattice monolayers decorated with OCs are stacked with AFM spin configuration. Physical properties such as polarization (P), magnetization (M), and MOKE ($\theta_{\text{K}}$) are activated and controllable by the twist. (d) TBL CrI$_3$ with 21.79$^\circ$ twist, where the CrI$_3$ octahedral surface coloring distinguishes the top and bottom layers by dark blue and sky blue, respectively.}
\label{fig:OCandTBL}
\end{figure}

Here, we demonstrate that MOKE is activated in AFM CrI$_3$ homo-bilayers by introducing a twist between the two monolayers. For commensurate moir\'e patterns, only certain twist angles give rise to MOKE activity, whereas others do not. The key to rationalize this counterintuitive result is the fact that CrI$_6$ octahedra have an alternating handedness [octahedral chirality -- OC; See Fig.~\ref{fig:OCandTBL} (a) and (b)], thus defining a staggered sublattice degree of freedom of the honeycomb lattice~\cite{gibertini_magnetism_2021, kong_switching_2021, yu_interlayer_2021, ghosh_chirality_2022}. Therefore, the interplay of twisting and octahedral chirality can lower the symmetry of the untwisted bilayer, thus allowing the presence of Kerr rotation. Remarkably, for the same twist angles also a non-zero electrical polarization results, thus making the twisted bilayer (TBL) multiferroic. The signs of the Kerr rotation and the polarization change by reversing the twist direction. This is conceptually depicted in Fig.~\ref{fig:OCandTBL} (c). Finally, we study the $k$-space spin textures of twisted AFM CrI$_3$ homo-bilayers and show that they display exotic topologies such as antichiral patterns of vortices or alternating spin sinks and sources. Our results demonstrate that the interplay between twisting and octahedral chirality in magnetic bilayers represents an exciting new platform for tailoring their physical and functional properties.

\section{Structures and Symmetries}
\label{sec:str_sym}
The crystal structure of bulk CrI$_3$ consists of honeycomb layers arising from a 2D Cr-I-Cr covalent bond network.  
In this case, the basic bonding topology can be represented as a honeycomb (defective triangular) lattice of edge-sharing MX$_6$ octahedra (M=metal, X=anion), in which one out of the three triangular sites is vacant (V). This arrangement is found in a variety of materials, including oxides (\textit{e.g.}, Na$_2$IrO$_3$~\cite{shitade_quantum_2009}) and sulfides (\textit{e.g.}, MnPS$_3$~\cite{kuo_exfoliation_2016}). Opposing triangular faces of each octahedron `bend' with respect to each other in order to fulfill the steric requirement of different V-X and M-X distances, while the triangular faces of adjacent octahedra are bent in opposite directions. This particular bonding network is typically characterized by alternating OCs.
In our case, the OC is defined at each Cr site by the connectivity to the surrounding six I ions and three nearest neighbor Cr ions. In particular, the Cr and the two I atoms bonded to it (a bidentate ligand) define a plane. A complex containing three bidentate ligands can take on the shape of a propeller: each Cr-2I-Cr plaquette is a `blade,' and the cluster becomes a `propeller' with three blades. The possible two configurations define the left (L) and right (R) OC as shown in Fig.~\ref{fig:OCandTBL} (a) and (b) \cite{NomenclatureInorgChem_Ch9, footnote1_OC}. The L and R OC are mirror images of each other and can not be superimposed on each other, defining a pair of enantiomers. Monolayer CrI$_3$ can therefore be considered as a honeycomb lattice `decorated' with alternating OC on the same layer, thus defining two sublattices, denoted as `L' and `R', \textit{i.e.}, a racemic mixture of R and L OCs.

\begin{figure}[t!]
\centering
\includegraphics[width=0.5\textwidth]{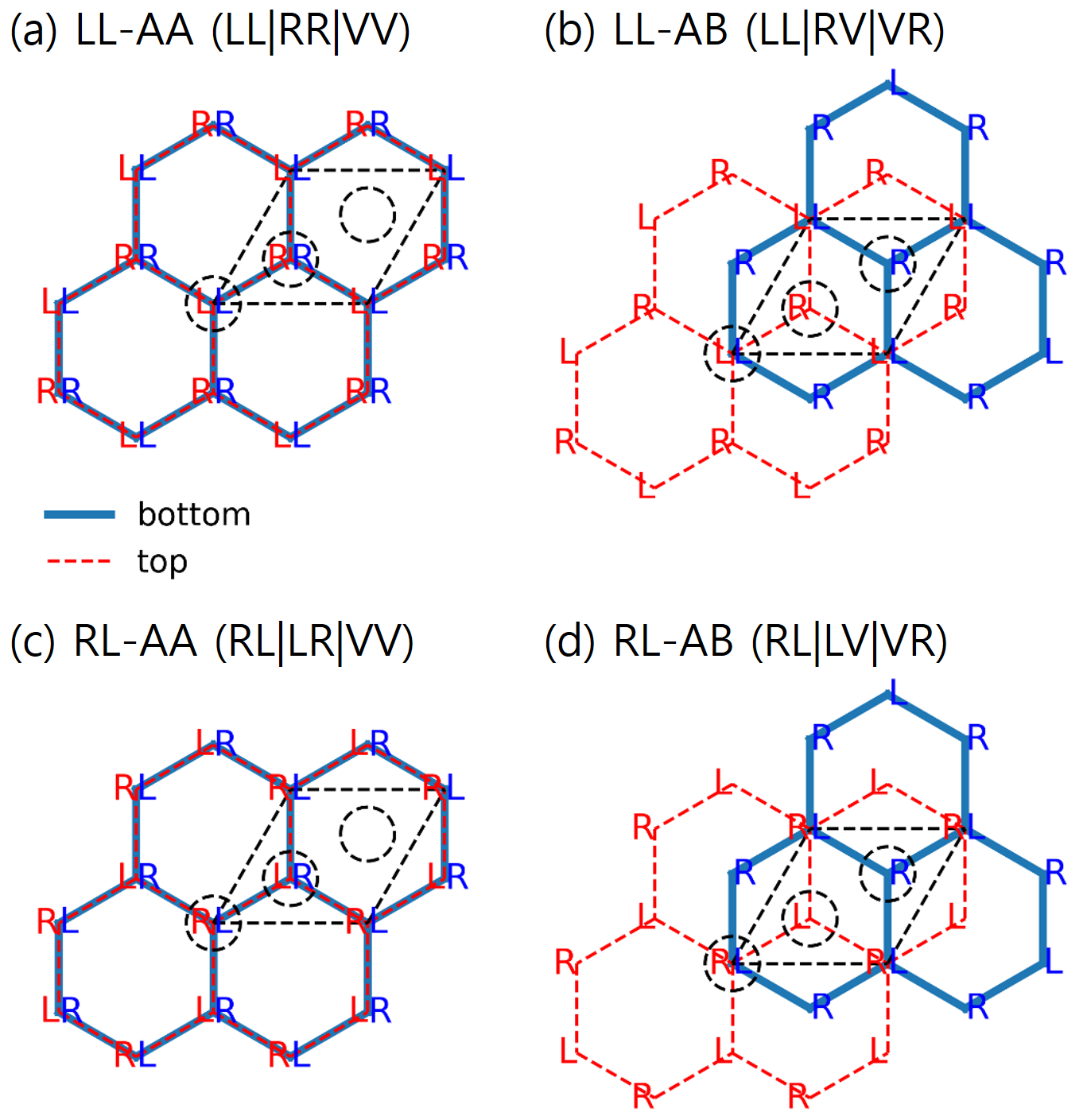}
\caption{Schematic pictures of the untwisted honeycomb lattice stackings, the LL family of (a) AA-stacking and (b) AB-stacking, and the RL family of (c) AA-stacking and (d) AB-stacking, are shown. 
Red and blue lines represent the honeycomb lattice of the top and bottom layers, respectively, with the OC labels (L and R) in the corresponding colors. Black solid lines and dashed circles are unitcells and 3-fold axis sites, respectively. If there is no red (blue) OC label in a circle, the corresponding label for the top (bottom) layer is V.
}
\label{fig:untwisted}
\end{figure}

In the bulk and in few-layer CrI$_3$ structures, CrI$_3$ monolayers are stacked on top of each other and may also be twisted. In this paper, we consider a particular set of CrI$_3$ TBLs that can all be obtained from exactly stacked monolayers (so-called AA stacking) by twisting around an orthogonal axis passing through a pair of stacked CrI$_6$ octahedra, the twist angle being such as to produce a commensurate supercell. Although this set does not exhaust all possible commensurate TBLs, it does include all cases that have at least 3-fold symmetry, and, as we shall see, it is sufficiently rich to produce a variety of behaviors in the macroscopic properties we want to explore. As a preliminary consideration, we note straight away that our structures set comprises two `families', depending on whether the CrI$_6$ octahedra stacked along the twist axis have the same or opposite OC.  We label these two families `LL' and `RL' respectively, with the understanding that each family also contains the structures obtained by global inversion symmetry (so, for example, the LL family also contains the RR-stacked structures). Before proceeding further with the symmetry analysis, we note that any 3-fold-symmetric 2D lattice contains exactly three 3-fold axes per unit cell (of which exactly one may be also a 6-fold axis), and that in each CrI$_3$ monolayer these axes must necessarily correspond either to a CrI$_6$ octahedron or to a vacancy. For example, in low-temperature bulk rhombohedral CrI$_3$ (space group $R\bar{3}$)~\cite{mcguire_coupling_2015} the stacking of each pair of monolayers is RL|VR|LV, where we denoted the three 3-fold positions by the local `stack' (first/second symbol corresponding to the top/bottom layer) and separated them by a vertical bar (|). It follows immediately that structures of the LL family must have exact stacking LL|RR|VV or LL|RV|VR (and inversion-related configurations), while those of the RL family must have exact stacking RL|LR|VV or RL|LV|VR~\cite{b_footnote2_3foldsites}. These untwisted stackings are schematically depicted in Fig.~\ref{fig:untwisted}.

We now proceed to consider the effects of different twisting angles on our two families, \textit{i.e.}, the interplay of twist and OC. We will proceed in two steps: first, we consider the symmetries obtained by twisting two AA-stacked honeycomb lattices around the AA-stacked sites; in a second step, we will consider the effect of OC. Since we will limit our description to commensurate superstructures, we will describe the resulting configurations using the so-called layer groups \cite{ITCvolE_2010}, which describe all structures that are periodic in two dimensions and have a finite extension in the third.

The case of bare honeycomb bilayers without any sublattice degrees of freedom is analogous to well-known TBL graphene. Only two layer-group symmetries are possible for commensurate supercells, $p622$ or $p321$, which we will refer to hereafter as the `T1 series' and the `T2 series', respectively. The T1 series has 6-fold axes, which must correspond to two stacked vacancies (VV) -- a configuration that never occurs in the T2 series. As already stated, there are always three sites in the supercell having 3-fold or 6-fold symmetry, conventionally set at (0, 0), (2/3, 1/3), and (1/3, 2/3) in fractional coordinates, all located along the long diagonal of the unit cell. It is also found that if a twist angle $\theta$ produces a (commensurate) T1 structure, then the twist angle (60$^{\circ}-\theta$) produces a T2 structure. Finally, we observe that the `untwisted' AA-stacked honeycomb bilayer is the $\theta=0$ member of the T1 series. Correspondingly, $\theta=60^{\circ}$ (T2 series) is the usual AB-stacked bilayer, which can also be obtained by translation. These two structures have additional inversion centers, thus increasing the symmetries to $p6/mmm$ (AA) and $p\bar31m$ (AB), respectively. 

\begin{figure*}[ht!]
\centering
\includegraphics[width=0.8\textwidth]{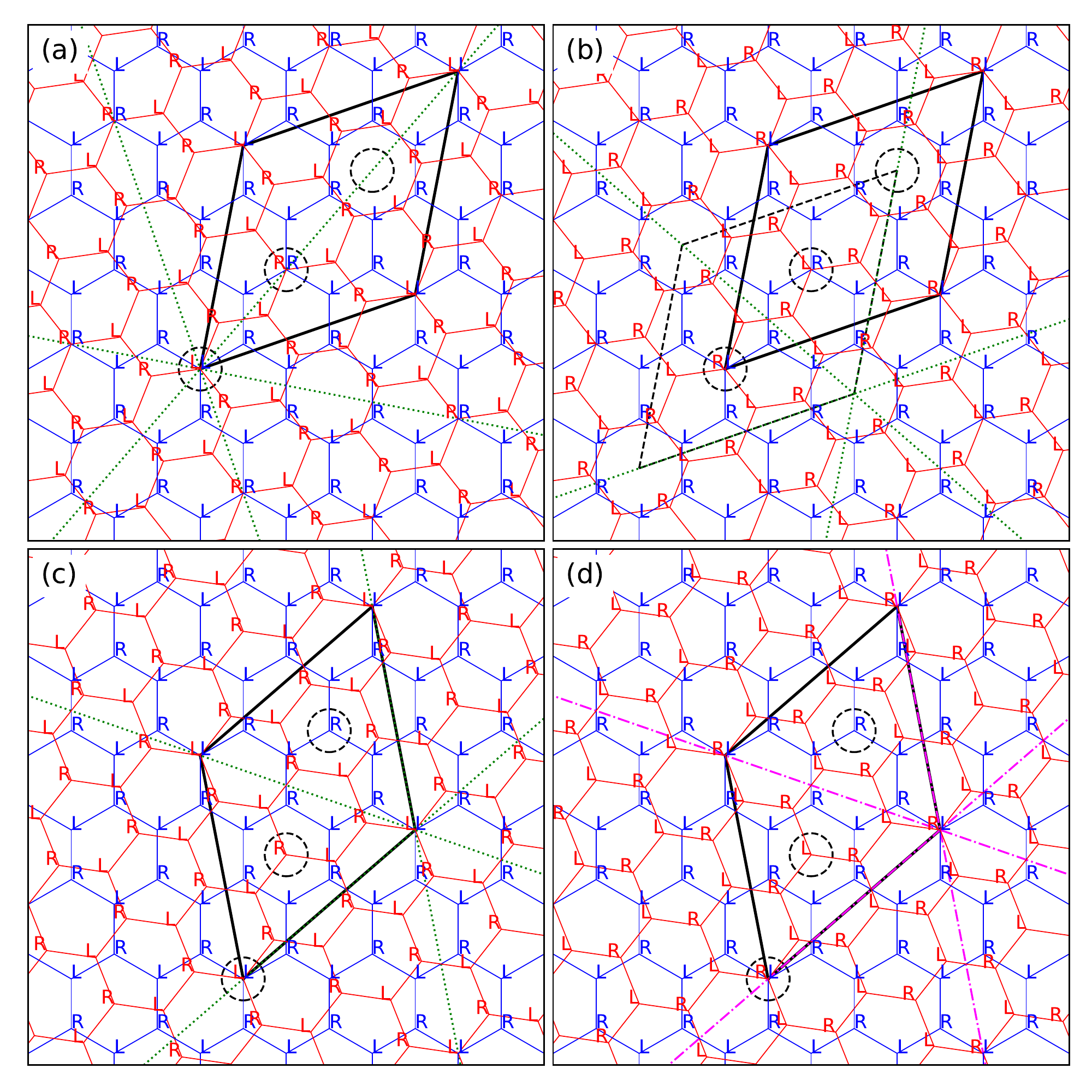}
\caption{
Schematic pictures of supercell, OC, and symmetry of TBL CrI$_3$.
Black solid lines are the supercell. Red and blue solid lines represent the honeycomb lattice of the top and bottom layers, respectively. L and R are labels for the left and right OC.
Black dashed circles denote the 3-fold rotation centers. Green dotted lines mean the 2-fold axis, and purple dash-dotted lines mean the lost 2-fold axis.
The systems with a  twist angle of 38.21$^\circ$ (T1) (a) of LL bilayer ($p312$) and (b) of RL bilayer ($p321$); 21.79$^\circ$ twist (T2) (c) of LL ($p321$) and (d) of RL bilayer ($p3$).}
\label{fig:symmetry1}
\end{figure*}

As the second step, we now decorate the honeycomb lattices with octahedra having alternating OC, and consider the combined effect of twist and OC, which results in further symmetry lowering. As explained above and in Appendix \ref{sec:T1T2}, each `series' (T1 and T2) will split in two distinct families (LL and RL). The T1 series gives rises to LL|RR|VV (LL family) and RL|LR|VV (RL family, both with the bare-honeycomb 6-fold axes at VV), while the T2 series produces LL|RV|VR (LL family) and RL|LV|VR (RL family), neither of them possessing a bare-honeycomb 6-fold axis. The layer-group symmetries of these structures are listed in Table~\ref{tab:symmetry}. Examples of the different situations and corresponding symmetries are shown schematically in Fig.~\ref{fig:symmetry1}, where we report projected views of the supercells with twist angles of 38.21$^\circ$ [T1 twist series, panels (a) LL and (b) RL] and 21.79$^\circ$ [T2, (c) LL and (d) RL].

{\renewcommand{\arraystretch}{1.2}
\begin{table}[b]
\centering
\begin{tabular}{c|l|l}
\hline
& LL & RL \\
\hline
\hline
 T1  & $p312$ (untwisted $p\bar{3}1m$) & $p321$ (untwisted $p\bar{6}2m$)   \\
 T2  & $p321$ (untwisted $p321$) & $p3$ (untwisted $p\bar{3}$) \\
\hline
\end{tabular}
\caption{\label{tab:symmetry}
Layer-group symmetries of TBL CrI$_3$. The `untwisted' structures correspond to AA stacking for the T1 series and AB stacking for the T2 series (see text).}
\end{table}}

Clearly, the interplay between twist and OC results in further symmetry lowering -- for example, the 6-fold symmetry of the T1 series is lost since it would connect octahedra with different OC. One interesting result is that the symmetry of the RL-T2 series is a \emph{polar} group $p3$, which allows out-of-plane electric polarization. At first sight, this may appear counterintuitive since the two monolayers, which are identical by symmetry in the untwisted structure, have to become inequivalent upon twisting.  
In fact, this result can be explained as a consequence of the interplay between OC and twist chirality. One can note that the twist itself induces global chirality~\cite{kim_chiral_2016, yananose_chirality-induced_2021} (right- and left-twists forming an enantiomeric pair) -- in other words, improper rotations, such as mirror, inversion, and roto-inversion, \emph{must} be absent in the twisted system. Consequently, the only symmetry operators that could connect the two monolayers are 2-fold rotations with horizontal axes.  It turns out that the presence or absence of two-fold axes is a characteristic of the series, including the untwisted end members. For example, the bulk rhombohedral structure ($R\bar{3}$) lacks 2-fold symmetry -- a characteristic that, at the bilayer level, is shared with the whole RL-T2 series.  We therefore conclude that the twisted members of the RL-T2 series lack any symmetry operator that can connect the two monolayers, which must therefore necessarily be inequivalent, leading to a polar (pyroelectric) layer group.  
Notably, this polarization mechanism based on global symmetry is distinct from that in TBL hexagonal boron nitride (hBN) which is based on the local polar structures~\cite{yasuda_stacking-engineered_2021, d_footnote4_hBN}.

\begin{figure*}[ht!]
\centering
\includegraphics[width=0.8\textwidth]{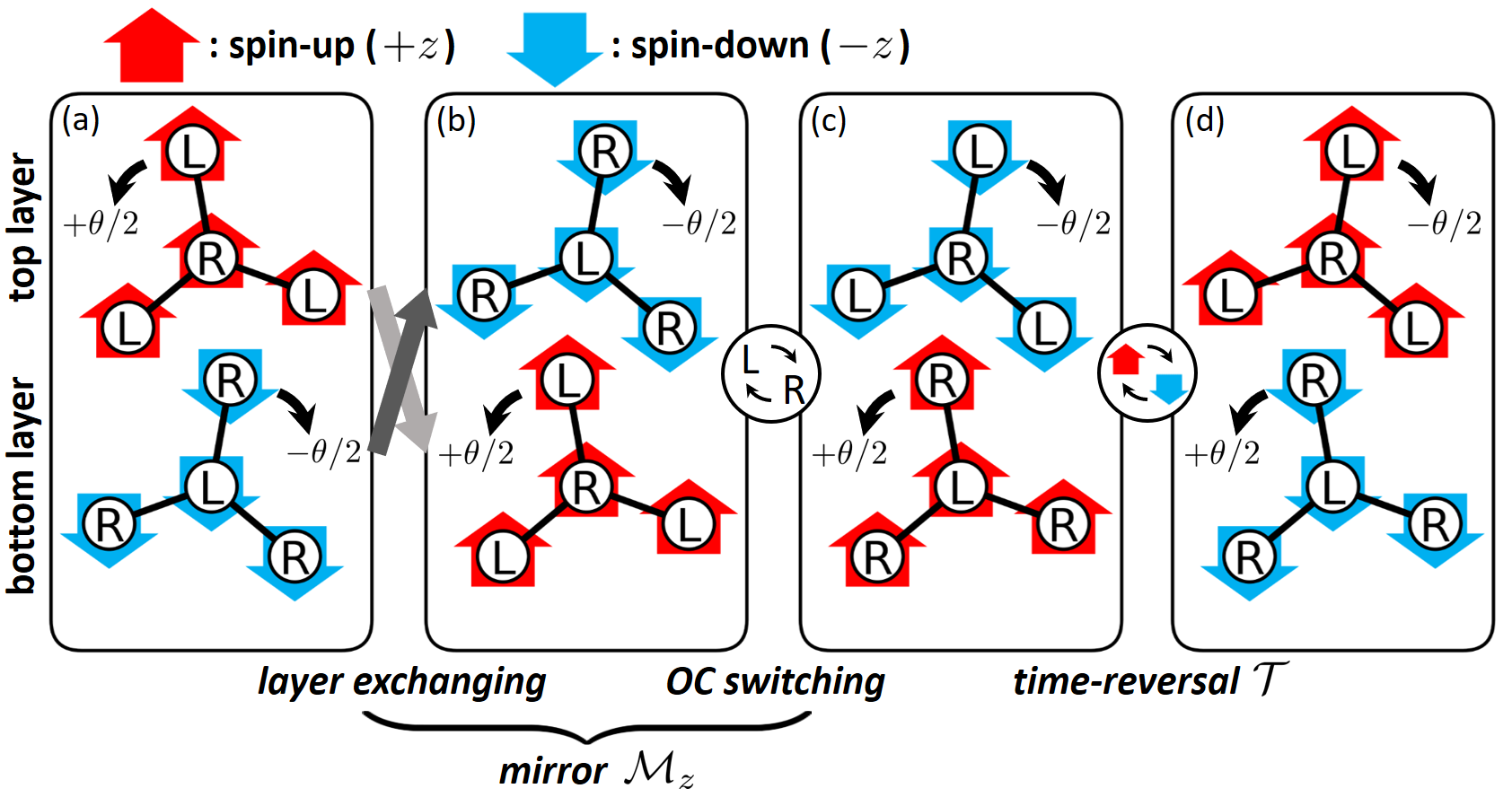}
\caption{
Schematic pictures depicting the effects of $\mathcal{M}_z$ and $\mathcal{T}\mathcal{M}_z$ on the twisted RL family. In each panel, the top and bottom layers are depicted with OCs, twist directions, and spin configurations. (a)$\rightarrow$(b) and (b)$\rightarrow$(c) are each of the two-step procedure of the mirror operation $\mathcal{M}_z$ representing the layer exchanging and OC switching, respectively. Note that the mirror operation leaves the spin component perpendicular to the mirror plane invariant, whereas the spin components parallel to the mirror plane are inverted. (c)$\rightarrow$(d) represents the time-reversal $\mathcal{T}$. Finally, (a) and (d) differ only by the sign of the twist angle $\theta$.}
\label{fig:TMz}
\end{figure*}

At the microscopic level, the monolayer inequivalence stems from the fact that the R and L octahedra will have slightly different bond lengths and angles in a twisted environment [See Appendix~\ref{sec:inequiv_stru}]. For example, the RL-T2 right-twisted structure has stackings RL|LV|VR at the 3-fold sites [See Fig.~\ref{fig:symmetry1} (d)]. 
From the previous discussions, it follows that the LV and VR stacks will be different due to different local structural properties and cannot be connected by any symmetry operation [See Appendices~\ref{sec:T1T2} and \ref{sec:inequiv_stru}]. 

As the third and final step, we need to consider the magnetic ordering and associated magnetic point-group symmetry. We assume that the magnetic ordering of TBL CrI$_3$ is AFM such that spins are coupled ferromagnetically within each layer but antiferromagnetically between the two layers, as shown by experimentally for the untwisted bilayer~\cite{huang_layer-dependent_2017}. 
Even though there are controversies on the true magnetic ground state and its dependence on bilayer stacking ~\cite{sivadas_stacking-dependent_2018, jang_microscopic_2019, soriano_magnetic_2020, sarkar_magnetic_2021, xu_coexisting_2022}, we consider here the AFM configuration, which is very suitable for highlighting our desired properties.
When the $p321$ or $p312$ layer group of TBL CrI$_3$ is combined with the AFM order, one obtains the magnetic point group $32$. On the other hand, layer group $p3$ with AFM corresponds to magnetic point group $3$, which is pyromagnetic (see below).

Let us now consider how the magnetic point-group symmetry determines the MOKE activity.
We consider polar-MOKE geometry, where both the light propagation and the magnetization axis of a sample are along the $z$-direction. Kerr rotation derives from the frequency-dependent optical conductivity tensor $\boldsymbol\sigma(\omega)$ according to the following formula~\cite{oppeneer_ab_1992, sangalli_pseudopotential-based_2012}.
\begin{equation}
\label{eq:MOKE_def}
    \theta_{\text{K}}+i\eta_{\text{K}} = \frac{-\sigma_{xy}}{\sigma_{xx}\sqrt{1+\tfrac{4\pi i}{\omega}\sigma_{xx}}}
\end{equation}
where $\theta_{\text{K}}$ is the Kerr rotation angle and $\eta_{\text{K}}$ is the Kerr ellipticity.
Thus a non-zero off-diagonal ($xy$) component of the conductivity tensor, or equivalently of the dielectric tensor according to the relation $\varepsilon_{\alpha\beta}=\delta_{\alpha\beta}+\tfrac{4\pi i}{\omega}\sigma_{\alpha\beta}$, implies a finite Kerr rotation.  
The 31 pyromagnetic point groups which allow for the presence of a finite magnetic moment can have finite off-diagonal components of conductivity tensor, and therefore such point groups are also MOKE active, even if the actual moment may be negligibly small~\cite{yang_magneto-optical_2020}.
For example, Yang, \textit{et al.}~\cite{yang_magneto-optical_2020} demonstrated that the AFM hetero-bilayer CrI$_3$/CrBr$_3$ induces non-zero Kerr rotation even in the vanishing net magnetic moment since it has the pyromagnetic group $3$.

We now consider twisted AFM homo-bilayer CrI$_3$.
For the $p321$ and $p312$ TBLs, which both share the  magnetic point group $32$,  the allowed form of the conductivity tensor is
\begin{equation}
\begin{split}
\boldsymbol{\sigma} = 
\begin{pmatrix}
 \sigma_{xx}  &  0  &  0  \\
 0  &  \sigma_{xx}  &  0   \\
 0  &  0  & \sigma_{zz}   \\
\end{pmatrix}.
\end{split}
\end{equation}
This implies that MOKE is  not active.
In contrast, the magnetic point group $3$ of the $p3$ TBL allows the conductivity tensor in the form of
\begin{equation}
\begin{split}
\boldsymbol{\sigma} = 
\begin{pmatrix}
 \sigma_{xx}  & \sigma_{xy} &  0  \\
 -\sigma_{xy} &  \sigma_{xx}  &  0   \\
 0  &  0  & \sigma_{zz}   \\
\end{pmatrix}.
\end{split}
\end{equation}
Since the off-diagonal component $\sigma_{xy}$ can be finite, the system may now become MOKE active~\cite{gallego_automatic_2019, yang_magneto-optical_2020}.

It is interesting to consider how MOKE activity is related in the TBL enantiomeric pairs.
For the RL family, regardless of T1 and T2 series, the structure obtained by the opposite twist angle ($-\theta$) is a mirror image of the original structure ($+\theta$), forming an enantiomeric pair. We can consider the mirror operation where the mirror plane is the middle plane of two layers ($\mathcal{M}_z$). The mirror operation  therefore exchanges two layers. This results in a sign change of twisting angle, $+\theta\rightarrow-\theta$, \textit{i.e.}, the switching of the twist chirality. In addition, the mirror operation changes the OC from L to R and viceversa. Therefore, one can examine how $\mathcal{M}_z$ acts by a two-step procedure. First, RL changes to LR by the layer exchanging, and then it becomes RL again by OC switching. Thus, the net effect of the $\mathcal{M}_z$ mirror is to produce a structure with the same layer-group symmetry and a $-\theta$ twist [See Fig.~\ref{fig:TMz} for schematic understanding and Appendix~\ref{sec:cell_generating} for a mathematical derivation]~\cite{c_footnote3_inv}.
If the magnetic ordering is considered, one has to introduce the $\mathcal{T}\mathcal{M}_z$ symmetry operation, where $\mathcal{T}$ is time-reversal, in order to generate the $-\theta$ structure.
Moreover, the dielectric tensor is transformed as  $(\mathcal{T}\mathcal{M}_z)\boldsymbol{\varepsilon}(\mathcal{T}\mathcal{M}_z)^{-1} = \boldsymbol{\varepsilon}^{T}$ with respect to that of $+\theta$ system (invariant under $\mathcal{M}_z$ and transpose by $\mathcal{T}$ according to Onsager's relation~\cite{rathgen_polar_2005}). It leads to a fascinating property for the $p3$ case (RL-T2): the sign of $\varepsilon_{xy}$ ($\sigma_{xy}$) is changed, and, therefore, one can reverse the Kerr angle (\textit{i.e.} sign change) by switching the twist chirality.
In addition, the $\mathcal{M}_z$ or $\mathcal{T}\mathcal{M}_z$ relation between the two enantiomeric pairs implies that the polarization perpendicular to the layer is inverted by twist chirality change (note that the polarization is invariant under $\mathcal{T}$).

We can draw our first conclusions from these considerations. According to symmetry analysis, one should expect electric polarization as well as MOKE activity in the AFM twisted CrI$_{3}$ homo-bilayers with layer-group symmetry $p3$, which originate from twisting the RL-stacked bilayers through the T2 angle series. By contrast, twists through the T1 series or T2 twists of the LL stacks should result neither in electrical polarization nor in MOKE activity. 
Note that, although the symmetry arguments for the activation and switching of the physical properties in the $p3$ system were made for collinear spin configurations, they would clearly remain valid for any non-collinear components sharing the same magnetic symmetry. On the other hand, non-collinear components (spin canting) breaking additional symmetries could induce polarization or MOKE even when they are absent in the collinear structures, though the behavior upon $+\theta\rightarrow-\theta$ switching would in general be different from the one described above. 
To confirm the ``activation'' of the physical properties by twisting, we performed first-principles density functional theory (DFT) and tight-binding (TB) model calculations. Specifically, we focus on the case of the twist by 21.79$^\circ$ (T2). For comparison, we also investigate the $p321$ system with the same 21.79$^\circ$ twist but of the LL bilayer, which, in contrast, is expected not to be active for the same phenomena.

\section{Methods}
Symmetries of the systems were identified by using both FINDSYM~\cite{stokes_findsym_2005} and Spglib~\cite{togo_spglib_2018} before and after atomic relaxations. The threshold on the symmetry check has been set to $2.5\times10^{-4}$ of \textit{atomic position tolerance} in FINDSYM.
To identify the electric conductivity tensor components allowed by a specific magnetic point group, the MTENSOR module of the Bilbao Crystallographic Server was used~\cite{gallego_automatic_2019}. 
The primitive cell of the monolayer CrI$_3$ used to generate the supercell has been optimized both in atomic positions and lattice constants. The optimized lattice constant $a_0 = 6.985$ \AA~ is in good agreement with the experimentally measured bulk value, $6.867$ \AA~\cite{mcguire_coupling_2015}.
We consider supercells with $\sqrt{7}\times\sqrt{7}$ periodicity resulting in a lattice constant $a = 18.481$ \AA. The out-of-plane direction lattice constant is $c = 24$ \AA~corresponding to the vacuum layer of nearly 10 \AA~in order to avoid the unphysical interactions between the periodic copies along the out-of-plane direction. 
In our settings, the magnetic moment of the top (bottom) layer points downward (upward).

For the DFT calculations, the Vienna Ab-initio Simulation Package (VASP)~\cite{kresse_efficient_1996} was used with projector augmented wave type pseudo potentials~\cite{kresse_ultrasoft_1999} and the  GGA-PBE exchange-correlation functional~\cite{perdew_generalized_1996}. $3\times 3\times 1$ regular $k$-point grid including $\Gamma$ point is used. Spin-orbit coupling (SOC) and on-site Coulomb repulsion correction by $U = 3$ eV and $J = 0.9$ eV are considered. The plane-wave basis energy cut-off is chosen to be 450 eV. The total energy convergence criterion was set as 10$^{-7}$ eV and the structure optimization was done with a criterion 0.001 eV/\AA. In order to calculate the electric polarization, we used the Berry phase method~\cite{king-smith_theory_1993}. To evaluate the electric polarization density of the 2D material, we adopted the thickness of the bilayer as $t = 2c_B/3$ where $c_B = 19.807$ \AA~is the out-of-plane direction lattice constant of rhombohedral bulk CrI$_3$~\cite{mcguire_coupling_2015} as done in Ref.~\cite{bruyer_possibility_2016}.
In the case of the out-of-plane polarization of a 2D material, there is no ambiguity from the polarization quanta~\cite{king-smith_theory_1993}. Thus there is no need to design a path in the configuration space in order to estimate the value of the polarization. 

An accurate Kerr angle calculation by the DFT requires a very high computational cost. On the other hand, Kerr angles of AFM TBL systems are expected to be very small. These facts make it difficult to compare the effect from the different symmetries of TBLs by DFT. To circumvent these limitations, we used the TB theory approach for calculating the MOKE.
First, we constructed the TB Hamiltonian of the monolayer. The $d$-orbitals are assigned as a basis set for Cr and $p$-orbitals for I, defining a total of 56 bands. Slater-Koster parametrization~\cite{slater_simplified_1954} and the symmetry-considered hopping parameters are used to construct the Hamiltonian. SOC is considered by the intra-atomic $\lambda\mathbf{L}\cdot\mathbf{S}$ perturbation terms~\cite{fang_ab_2015}.
The hopping parameters were extracted from the Wannier interpolated Hamiltonian~\cite{marzari_maximally_1997, franchini_maximally_2012} obtained from the DFT calculation for the monolayer. 
Then, the hopping parameters from the monolayer are used to construct the Hamiltonian of the TBLs.
Inter-layer interactions are introduced as the Slater-Koster type hopping between the upper I's of the bottom layer and the lower I's of the top layer. By adopting the exponentially decaying hopping strength, arbitrary inter-layer geometries given by the TBL structures can be considered~\cite{trambly_de_laissardiere_localization_2010, fang_ab_2015}.
Note that the construction of the TB model in addition to the Wannierization allows us to implement all symmetry restrictions on our systems.
Further details of the TB model are described in Sec. SI of Supplemental Material (SM)~\cite{Suppl}.
The polarization is calculated also by the TB model~\cite{solovyev_magnetic-field_2010, barone_ferroelectricity_2011}, as well as by the DFT. 

MOKE spectra were calculated based on Eq.~(\ref{eq:MOKE_def}).
Within the Independent Particle Random-Phase Approximation (IP-RPA) scheme,
conductivity tensors are calculated from the Kubo-Greenwood formula~\cite{oppeneer_ab_1992, wang_band_1974, yao_first_2004, fang_large_2018, kumar_gudelli_magnetism_2019},
\begin{equation}
\sigma_{\alpha\alpha}(\omega) = \frac{-i\hbar e^2}{N_{k}V}\sum_{\mathbf{k}nm}
\frac{f_{\mathbf{k}m}-f_{\mathbf{k}n}}{E_{\mathbf{k}m}-E_{\mathbf{k}n}}
\frac{|(\mathbf{v}_{mn}^{\mathbf{k}})_{\alpha}|^2}
{\hbar\omega+E_{\mathbf{k}m}-E_{\mathbf{k}n}+i\eta}
\end{equation}
for the diagonal components and 
\begin{equation}
\sigma_{\alpha\beta}(\omega) = \frac{\hbar e^2}{N_{k}V}\sum_{\mathbf{k}nm}
(f_{\mathbf{k}m}-f_{\mathbf{k}n})
\frac{\Im(\mathbf{v}_{mn}^{\mathbf{k}})_{\alpha}(\mathbf{v}_{nm}^{\mathbf{k}})_{\beta}}
{(E_{\mathbf{k}m}-E_{\mathbf{k}n})^2-(\hbar\omega+i\eta)^2}
\end{equation}
for the off-diagonal components.
$N_{k}$ is the number of $k$-point grid, which is $9\times 9$ in this work ($3\times 3$ for twist angle dependence in Fig.~\ref{fig:MOKEangle}). 
$V$ is the cell volume defined with the layer thickness. 
$f_{\mathbf{k}n}$ is the occupation number that is 1 for occupied and 0 for unoccupied states, \textit{i.e.}, insulator.
$E_{\mathbf{k}n}$ is energy eigenvalue, and  
$\eta$ is the broadening factor, in our case, 0.1 eV.
The velocity matrix $\mathbf{v}^{\mathbf{k}}_{nm} \equiv \mel{\psi^{\mathbf{k}}_{n}}{(i/\hbar)[\hat{H},\hat{\mathbf{x}}]}{\psi^{\mathbf{k}}_{m}} = \mel{u^{\mathbf{k}}_{n}}{(1/\hbar)\boldsymbol{\nabla}_{\mathbf{k}}\hat{H}_{\mathbf{k}}}{u^{\mathbf{k}}_{m}}$ is obtained by the effective TB velocity operator formalism~\cite{graf_electromagnetic_1995}, \textit{i.e.}, $(\mathbf{v}^{\mathbf{k}}_{nm})^{\text{TB}} = (1/\hbar)\mathbf{C}^{\dagger}_{\mathbf{k}n}(\boldsymbol{\nabla}_{\mathbf{k}}H_{\mathbf{k}})\mathbf{C}_{\mathbf{k}m}$ where $H_{\mathbf{k}}$ and $\mathbf{C}_{\mathbf{k}n}$ are the TB Hamiltonian matrix and eigenvectors.
For the validation of the monolayer TB Hamiltonian, the $\mathbf{v}^{\mathbf{k}}_{nm}$ obtained via the Wannier interpolation method is used as a DFT level result.

Since we are focusing on the role of symmetry, we limit our study to Eq.~(\ref{eq:MOKE_def}) with the IP-RPA method. This means that we ignore the effect of a substrate~\cite{fang_large_2018, wu_physical_2019, kumar_gudelli_magnetism_2019, molina-sanchez_magneto-optical_2020} or excitons~\cite{wu_physical_2019, molina-sanchez_magneto-optical_2020}.

\section{Results and Discussions}

\subsection{Structure, Polarization, and Magnetism}
We constructed the ideal systems by a rigid rotation of the perfect monolayers. The symmetry, after including the atomic relaxations in the twisted systems, remains unchanged within the fixed threshold. 
Electric polarizations obtained from DFT are listed in Table~\ref{tab:polandenergy}. 
As symmetry implies, the $p321$ system has no electric polarization before and after the relaxation. In contrast, the $p3$ system has a finite value of polarization. When considering the ideal systems, the ionic contribution to the polarization in the $p3$ system is equal to that of the $p321$ system since each monolayer is identical. This means that the polarization is purely electronic, which is allowed in a polar symmetry group. 
In addition, as suggested by previous symmetry arguments, our DFT calculations clearly confirmed that the $p3$ system with right- and left-twisting ($\pm21.79^\circ$) exhibit exactly opposite polarization, \textit{i.e.}, the change of twist chirality of the system inverts the polarization. 

In the TB model, the $p3$ right-twist system shows the polarization of $-0.234$ nC/cm$^2$, which has the same sign and order as the DFT result $-0.812$ nC/cm$^2$; moreover, it is inverted in the left-twist system. The difference between the DFT and TB values originates from that the TB model can not consider the charge redistribution in TBLs, as well as that the monolayer electronic structure of the TB model deviates from that of the DFT. 
As expected, the polarization is 0 in the $p321$ system. Therefore, the TB model results are qualitatively consistent with the DFT results.

\begin{figure}[t!]
\centering
\includegraphics[width=0.5\textwidth]{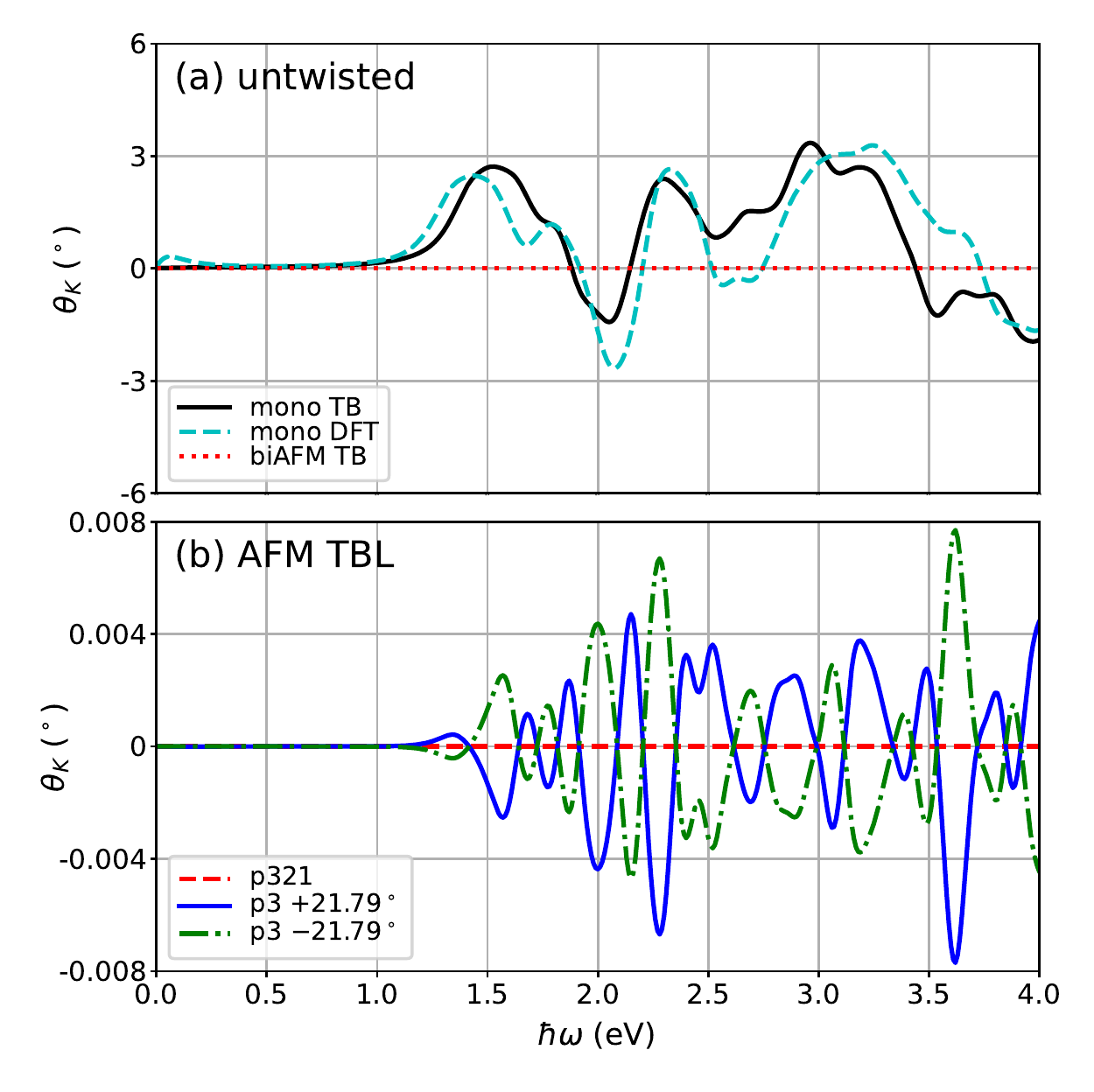}
\caption{
MOKE spectra in degrees with respect to the photon energy $\hbar\omega$ of
(a) FM monolayer (mono) and untwisted AFM bilayer (biAFM), and (b) AFM TBLs CrI$_3$. `TB' and `DFT' means the calculation methods, where DFT implies the Wannier interpolation method. MOKE spectra of TBLs are all calculated by TB. 
$p321$ and $p3$ are LL-T2 and RL-T2 systems, respectively.
The AFM bilayer and $p321$ system show vanishing spectra, and the two $p3$ systems exhibit exactly opposite spectra.}
\label{fig:MOKE}
\end{figure}

{\renewcommand{\arraystretch}{1.1}
\begin{table}[b]
\centering
\begin{tabular}{l|c|c}
\hline
system  & $P$ (nC/cm$^2$) & $E_{\text{AFM}} - E_{\text{FM}}$ (meV)\\
\hline
\hline
$p321$ (unrelaxed) & $\ \ 0.000$ & 0.1165 \\
$p321$ (relaxed) & $\ \ 0.000$ & 0.0265 \\
$p3$ (unrelaxed) & $-0.812$ & 0.1119 \\
$p3$ (relaxed) & $-0.057$ & 0.0259 \\
\hline
\end{tabular}
\caption{\label{tab:polandenergy}
Out-of-plane ($z$) direction of the electric polarization densities and the total energy difference per f.u. of the AFM and FM phases for the TBLs from DFT.
$p321$ and $p3$ are LL-T2 and RL-T2 systems, respectively.}
\end{table}}

We compared the total energies of FM and AFM phases in the same structures as reported in Table~\ref{tab:polandenergy}. The two phases are almost degenerate, with a slight preference for the FM phase. 
Indeed, a very recent experiment reported the FM phase for large angle TBL CrI$_3$~\cite{xu_coexisting_2022}, as opposed to the early observation of the AFM in the untwisted bilayer~\cite{huang_layer-dependent_2017}. Moreover, earlier DFT studies~\cite{sivadas_stacking-dependent_2018, jang_microscopic_2019} indicated that the untwisted bilayer can be either FM or AFM depending on the stacking. Thus, the discrepancy in the preferred magnetic ground state should be traced back to the details of the stacking.
Nevertheless, the energy difference between the two phases is very small, and the true ground state can depend on various factors including the experimental details. For the purpose of our work, we focus only on the homogeneous AFM phase, where the twist gives rise to the switchability of the Kerr rotation.
In order to obtain a single domain condition, a large-twist angle system might be suitable, whose supercell is small and thus incompatible with the occurrence of phase domains.

We calculated the magnetic moments from the TB model. Since the 3-fold symmetry does not allow any in-plane moment, all the in-plane components vanish, and only the $z$-component could be non-zero. The magnetic moment of the monolayer is $3.011\ \mu_B$ per formula unit (f.u., CrI$_3$), which is very close to the DFT result of $3.016\ \mu_B$. In the case of untwisted AB-stacked bilayer, the total magnetic moment vanishes, \textit{i.e.}, the TB model well reproduces the AFM configuration.  
In the $p3$ TBL system, the total out-of-plane moment, which is a result of inequivalence of two layers upon twisting, is very small, but not zero, \textit{i.e.}, $4.73\times10^{-8}\ \mu_B$ per f.u..  Moreover, the out-of-plane moment is inverted in the enantiomeric partner, which is consistent with the $\mathcal{T}\mathcal{M}_z$ relation. On the other hand, the moment vanishes in the $p321$ system, as expected. 
The very small magnitude in the $p3$ system is comparable with the limitation of the numerical accuracy in DFT simulations. This clearly suggests the necessity of investigations by the TB model in our cases.

\begin{figure}[t!]
\centering
\includegraphics[width=0.5\textwidth]{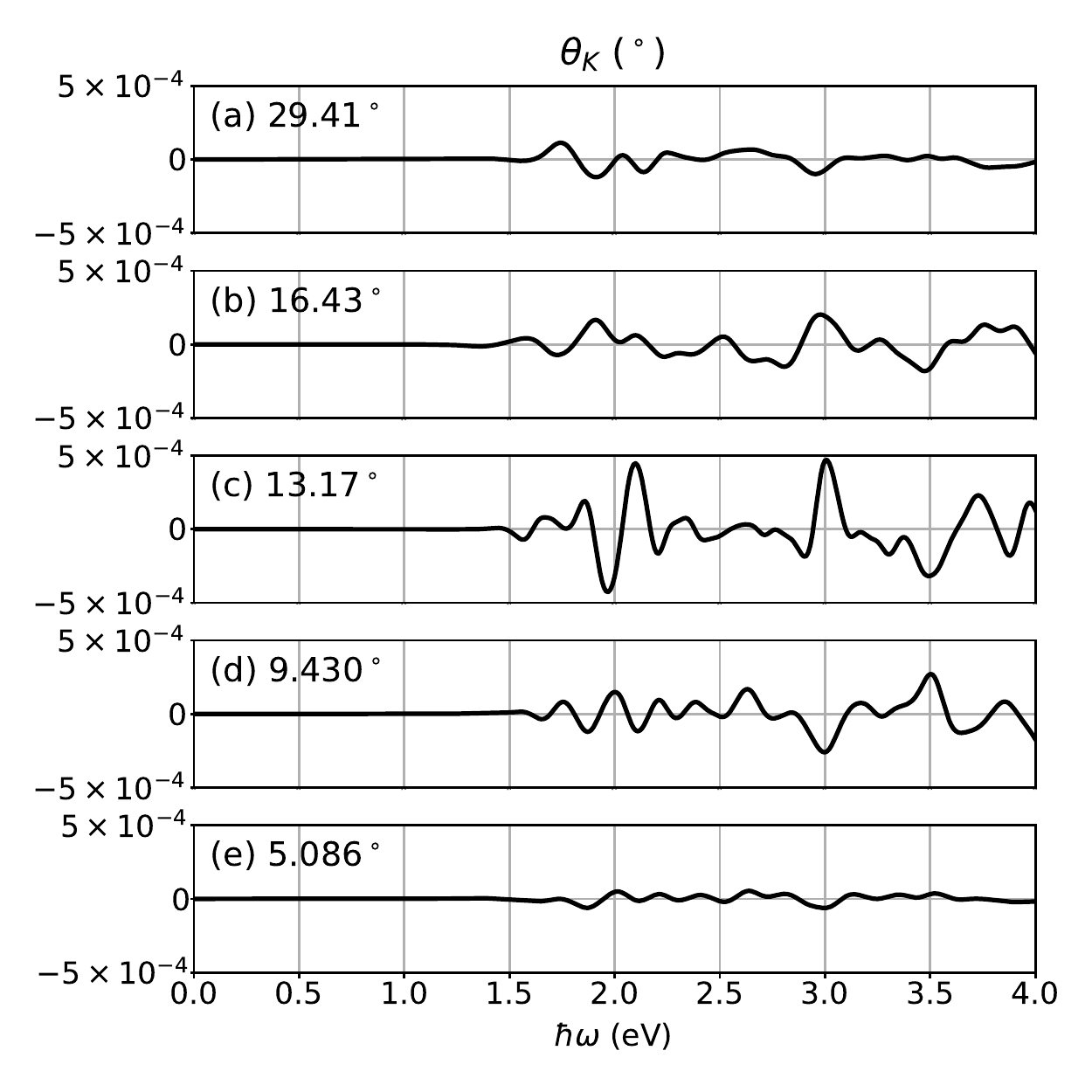}
\caption{
MOKE spectra of the $p3$ systems (RL-T2) with various twist angles from the TB model. (a) 29.41$^\circ$, (b) 16.43$^\circ$, (c) 13.17$^\circ$, (d) 9.430$^\circ$, and (e) 5.086$^\circ$. Each system exhibits different spectra.}
\label{fig:MOKEangle}
\end{figure}

\subsection{MOKE}

Before investigating the MOKE spectra of the TBL CrI$_3$, we calculated it for both the FM monolayer and the untwisted AB-stacked AFM bilayer, which are shown in Fig.~\ref{fig:MOKE} (a). For the monolayer, the Kerr angle obtained from the DFT using the Wannier interpolation method is also shown for comparison. The MOKE spectrum from the TB model reproduces well the main characteristics of the DFT result: a positive peak appears near 1.5 eV followed by a smaller bump, the next negative peak appears, again a positive peak follows it, and so on.  
On the other hand, the TB model reproduces a zero Kerr angle for the bilayer as expected from symmetry~\cite{huang_layer-dependent_2017, yang_magneto-optical_2020}.
Therefore, our TB model can be considered as a good starting point for investigating the MOKE of TBL systems, where direct DFT evaluation becomes exceedingly expensive in terms of the computational workload. 

We now proceed to discuss the spectra for TBLs. Fig.~\ref{fig:MOKE} (b) shows the calculated MOKE  of AFM TBLs. The $p321$ system does not show any MOKE signal. On the other hand, a small but finite Kerr angle appears in the $p3$ system, consistent with the previous symmetry arguments.
Moreover, the enantiomeric pair of the $p3$ systems, which have opposite twist angles, show exactly opposite MOKE spectra, as expected.
Note that the peak amplitude of the Kerr angle in the TBL system is $\sim0.008^\circ$, which appears rather small, but comparable to the values ranging from $\sim0.017^\circ$ to $\sim0.6^\circ$ obtained in literature~\cite{feng_large_2015, sivadas_gate-controllable_2016, higo_large_2018, zhou_spin-order_2019, yang_magneto-optical_2020, zhou_crystal_2021}.

We also calculated MOKE spectra for the $p3$ TBLs with different twist angles (Fig.~\ref{fig:MOKEangle}). The details of the systems and the calculated polarizations and magnetizations are listed in Table~\ref{tab:angles}, where we also include the $21.79^\circ$ case. All MOKE spectra almost vanish below 1 eV, which is consistent with the estimated magnitude of the energy gap (See SM Sec.~\textrm{SI} E),  but are different above this energy.
Interestingly, their amplitudes differ depending on the twist angle, although there is no clear dependence. On the other hand, there is a rough tendency that the larger the area of the TBL unit cell is, the smaller the amplitude; a similar tendency is found in the polarization: the larger the area, the smaller the polarization is.

{\renewcommand{\arraystretch}{1.1}
\begin{table}[b]
\centering
\begin{tabular}{C{0.04\textwidth}C{0.06\textwidth}C{0.04\textwidth}C{0.08\textwidth}C{0.11\textwidth}C{0.10\textwidth}}
\hline
($n$,$m$) & \begin{tabular}{@{}c@{}c@{}}Twist\\ Angle\\ ($^\circ$)\end{tabular} & $A/A_0$ & $L$ (\AA) & \begin{tabular}{@{}c@{}}$P$ \\ ($10^{-3}\ $nC/cm$^2$)\end{tabular} & \begin{tabular}{@{}c@{}}$m_z$ \\ ($10^{-9}\ \mu_B$)\end{tabular}\\
\hline
\hline
(8,3) & 29.41  & 97   & 68.7969  & $-2.72$ & $-0.412 $ \\
(2,1) & 21.79  & 7    & 18.4812  & $-234 $ & $47.3   $ \\
(5,3) & 16.43  & 49   & 48.8968  & $-1.91$ & $0.369  $ \\
(3,2) & 13.17  & 19   & 30.4480  & $-5.39$ & $-4.81  $ \\
(4,3) & 9.430  & 37   & 42.4896  & $-4.27$ & $5.55   $ \\
(7,6) & 5.086  & 127  & 78.7198  & $-0.72$ & $2.34   $ \\
\hline
\end{tabular}
\caption{\label{tab:angles}
TBL CrI$_3$ systems with various twist angles and their polarization and magnetization. ($n$,$m$) is a pair of integers that generate the TBL (See Appendix~\ref{sec:cell_generating}). $A/A_0$ is the relative area with respect to the area of the untwisted unit cell. $L$ is the TBL lattice parameter. $m_z$ is the total magnetic moment per f.u..}
\end{table}}

\subsection{MOKE Layer Decomposition}
\label{sec:layer_decomp}
In the case of the AFM bilayer, it is possible to consider that the two layers give rise to the opposite contributions to the MOKE spectra, which, however, do not cancel completely when the two layers are inequivalent. Therefore, we propose a layer-projected contribution analysis. This approach can provide further insights into the origin of the finite Kerr angle.
Starting from the TB Hamiltonian, the eigenvectors can be decomposed into a contribution from the top and bottom layer, \textit{i.e.}, $\mathbf{C}^{\dagger}_{\mathbf{k}n} = (\mathbf{c}^{\text{t}\dagger}_{\mathbf{k}n}\mathbf{c}^{\text{b}\dagger}_{\mathbf{k}n})$, where t and b denote the top and bottom layer. In the same way, the effective TB velocity operator can be decomposed as 
\begin{equation}
(\hat{\mathbf{v}}^{\mathbf{k}})^{\text{TB}} = 
\begin{pmatrix}
\mathbf{v}^{\text{t}\mathbf{k}} & \mathbf{v}^{\text{IL}\mathbf{k}} \\
\mathbf{v}^{\text{IL}\mathbf{k}\dagger} & \mathbf{v}^{\text{b}\mathbf{k}} \\
\end{pmatrix}
\end{equation}
where IL means the Inter-Layer contribution.
The matrix element can be decomposed accordingly.
\begin{equation}
\label{eq:v_decomp}
\begin{split}
(\mathbf{v}^{\mathbf{k}}_{nm})^{\text{TB}} &=
\mathbf{c}^{\text{t}\dagger}_{\mathbf{k}n}\mathbf{v}^{\text{t}\mathbf{k}}\mathbf{c}^{\text{t}}_{\mathbf{k}m} +
\mathbf{c}^{\text{b}\dagger}_{\mathbf{k}n}\mathbf{v}^{\text{b}\mathbf{k}}\mathbf{c}^{\text{b}}_{\mathbf{k}m} \\ 
&+
\mathbf{c}^{\text{t}\dagger}_{\mathbf{k}n}\mathbf{v}^{\text{IL}\mathbf{k}}\mathbf{c}^{\text{b}}_{\mathbf{k}m} +
\mathbf{c}^{\text{b}\dagger}_{\mathbf{k}n}\mathbf{v}^{\text{IL}\mathbf{k}\dagger}\mathbf{c}^{\text{t}}_{\mathbf{k}m} \\
&\equiv \mathbf{v}^{\text{t}\mathbf{k}}_{nm} + \mathbf{v}^{\text{b}\mathbf{k}}_{nm} + \mathbf{v}^{\text{IL}\mathbf{k}}_{nm}
\end{split}
\end{equation}
The off-diagonal component of the conductivity tensor $\sigma_{xy}$ is related to $(\mathbf{v}^{\mathbf{k}}_{mn})^{\text{TB}}_{x}(\mathbf{v}^{\mathbf{k}}_{nm})^{\text{TB}}_{y}$.
It gives rise to four contributions following Eq.~(\ref{eq:v_decomp}),
top layer term
$(\mathbf{v}^{\text{t}\mathbf{k}}_{mn})_{x}(\mathbf{v}^{\text{t}\mathbf{k}}_{nm})_{y}$,
bottom layer term
$(\mathbf{v}^{\text{b}\mathbf{k}}_{mn})_{x}(\mathbf{v}^{\text{b}\mathbf{k}}_{nm})_{y}$,
interlayer term
$(\mathbf{v}^{\text{IL}\mathbf{k}}_{mn})_{x}(\mathbf{v}^{\text{IL}\mathbf{k}}_{nm})_{y}$,
and cross-term components.
As a result, $\sigma_{xy}$ is represented  as follows.
\begin{equation}
\sigma_{xy} = \sigma_{xy}^{\text{t}} + \sigma_{xy}^{\text{b}} + \sigma_{xy}^{\text{IL}} + \sigma_{xy}^{\text{cross}}
\end{equation}
In this approach, by calculating the Kerr rotation from Eq.~(\ref{eq:MOKE_def}) but using, for instance, $\sigma_{xy}^{\text{t}}$ instead of total $\sigma_{xy}$, one can obtain the top layer contribution to the MOKE spectrum. For the $\sigma_{xx}$, the total value is used, so that the sum of each contribution is simply the total value. Note that the $\sigma_{xy}$ is only in the numerator of Eq.~(\ref{eq:MOKE_def}), thus permitting the sum decomposition.

\begin{figure}[t!]
\centering
\includegraphics[width=0.5\textwidth]{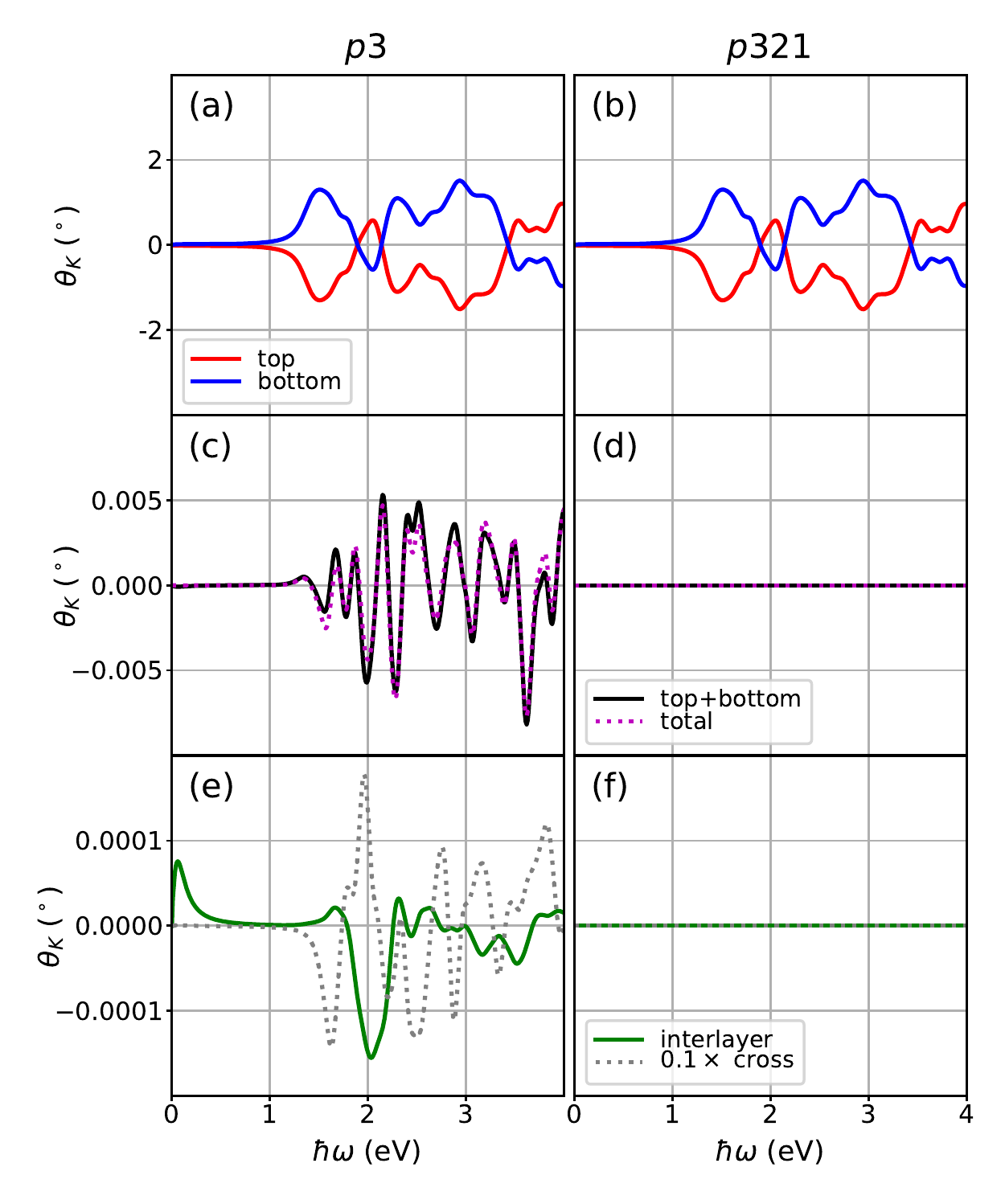}
\caption{
Layer-decomposed MOKE spectra of 21.79$^\circ$ $p3$ (RL-T2) and $p321$ (LL-T2) systems. (a,b) Top and bottom layer contributions to MOKE spectra calculated by the method introduced in Sec.~\ref{sec:layer_decomp}. (c,d) Comparisons between the summation of the top and bottom layer contributions shown in panels (a,b) and the total MOKE spectrum shown in Fig.~\ref{fig:MOKE} (b). (e,f) Interlayer and cross-term (on a 1/10 scale) contributions.}
\label{fig:MOKElayer}
\end{figure}

Fig.~\ref{fig:MOKElayer} shows the layer-decomposed MOKE spectra of $p3$ and $p321$ TBL systems.
The top and bottom layer contributions shown in Fig.~\ref{fig:MOKElayer} (a) and (b) are opposite in both cases, so the two contributions compensate. However, they do not cancel exactly for the $p3$ system.
In this case, the sum of the two layers' contributions shown in Fig.~\ref{fig:MOKElayer} (c) does not exactly cancel, and it almost overlaps with the total MOKE spectrum. This implies that the interlayer and the cross-term contributions have almost no effect on the finite MOKE.
In particular, the interlayer contribution is smaller than the total value [See Fig.~\ref{fig:MOKElayer} (e)] by two orders of magnitude. This suggests that the inequivalence between the top and bottom layers induced by symmetry lowering is the origin of the finite MOKE rather than the interlayer transitions.
The magnitude of the cross-term contribution is intermediate between the total value and the interlayer contribution [Fig.~\ref{fig:MOKElayer} (e)]. This can be inferred from the large velocity matrix elements from each layer and the small interlayer elements that are mixed in the cross-term.  
In the $p321$ system, instead, the top and bottom layer contributions are exactly canceled [Fig.~\ref{fig:MOKElayer} (d)] in agreement with symmetry considerations, and other contributions vanish [Fig.~\ref{fig:MOKElayer} (f)]. 
The untwisted AFM bilayer (not shown) also exhibits the exact cancellation between the two layers and vanishing interlayer and cross-term contributions like the $p321$ system, thus supporting the consistency of this approach.

\subsection{Spin textures in $k$-space}
\begin{figure}[t!]
\centering
\includegraphics[width=0.5\textwidth]{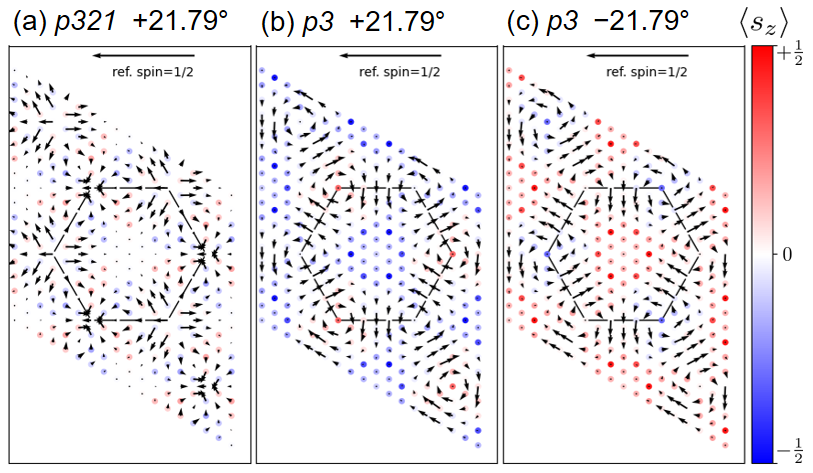}
\caption{
spin textures of the highest valence band of the AFM TBLs in the $k$-space. (a) $p321$ structure of $+21.79^\circ$ twist. $p3$ structure of (b) $+21.79^\circ$ and (c) $-21.79^\circ$ twist. Black solid lines represent the Brillouin zone boundary. In-plane spin components are shown by arrows, whose magnitude can be compared with the reference spin $1/2$ (eigenstate with spin $1/2$ along a given orientation) at the top of each panel. The color map represents the out-of-plane spin component. }
\label{fig:spintexture}
\end{figure}

It is interesting to consider the spin texture in $k$-space of the different systems. 
Spin textures are defined as $\mathbf{s}_{\mathbf{k}} = \mel{\psi_{\mathbf{k}}}{\mathbf{S}}{\psi_{\mathbf{k}}}$, where $\mathbf{S}$ is the spin operator, and they are shown in Fig.~\ref{fig:spintexture} for the highest valence bands of the $p321$ $+21.79^\circ$ [panel (a)] and the $p3$ $\pm 21.79^\circ$ [(b) and (c)] systems.
Interestingly, the $p3$ systems exhibit spin-texture vortices around the $K$ points, which are vertices of the Brillouin zone boundary. The \emph{local} chirality (whether the vortex whirls clockwise or counter-clockwise) is alternating for neighboring $K$ points, \textit{i.e.}, the spin texture is `antichiral'. The term antichiral has been recently introduced in the field of skyrmions~\cite{rybakov_antichiral_2021} while a few works have reported the antichiral spin textures in $k$-space without employing this terminology~\cite{loder_momentum-space_2017, farooq_spinvalley_2020, soriano_spinorbit_2021, ghosh_chirality_2022}. 
On the other hand, the $p321$ system exhibits radial spin textures of alternating `sink' and `source' around $K$ points.
The spin textures for different twist angles are shown in Sec. \textrm{SIII} of SM.

Clearly, the spin textures respect the symmetry of each system. In the $p321$ system, the spin texture has 3-fold symmetry around the $z$-axis and 2-fold symmetry around in-plane axes, while in the $p3$ system, it has only the 3-fold symmetry.
Moreover, the spin textures in two enantiomeric partners of the $p3$ satisfy the $\mathcal{T}\mathcal{M}_z$ relation.
The transformation rules for spin and $\mathbf{k}$ by each operation are $\mathcal{M}_z : (s_x,s_y,s_z)\mapsto(-s_x,-s_y,s_z)$ and $\mathbf{k}\mapsto\mathbf{k}$; and $\mathcal{T} : (s_x,s_y,s_z)\mapsto(-s_x,-s_y,-s_z)$ and $\mathbf{k}\mapsto-\mathbf{k}$.
As a consequence, when spin components of the $p3$ system with  $+\theta$ twist  at a specific $\mathbf{k}$ are given by $\mathbf{s}^{+\theta}_{\mathbf{k}} = (s_x,s_y,s_z)$, one can expect $\mathbf{s}^{-\theta}_{-\mathbf{k}} = (s_x,s_y,-s_z)$ for the $p3$ system with $-\theta$ twist [See Fig.~\ref{fig:spintexture} (b) and (c)]. This $s_z$ sign change correlates with the Kerr angle switching.

\section{Conclusions}
In this work, we considered the physical properties arising from the twisting of two magnetic monolayers. For a CrI$_3$ layer, the OC is defined for each Cr site as R or L, and this can be illustrated, for example, by the propeller-shaped geometry of Cr-2I-Cr neighboring bonds.
CrI$_3$ monolayers and untwisted bilayers are not globally chiral, because both OC enantiomers R and L are equally represented, thus forming a `racemic structure.' 
This situation changes radically in the TBLs: in fact, the twist of the two layers naturally introduces global chirality ~\cite{yananose_chirality-induced_2021}, which can interact with OC (a chiral sublattice degree of freedom) to reduce symmetry further in a variety of ways.
We have shown that commensurate CrI$_3$ TBLs have different layer-group symmetries depending on both OC stacking and the twist angle.  For example, for the $21.79^\circ$ twist angle, $p321$ and $p3$ symmetries are obtained for LL and RL stacking, respectively. Remarkably, the $p3$ structure is expected to have a net electric polarization in spite of the fact that the two layers are identical prior to twisting. Furthermore, symmetry dictates that MOKE activity must be present in $p3$ symmetry even for an AFM spin configuration. By contrast, the non-polar $p321$ structures show neither MOKE nor electric polarization. 

We verified these predictions by first-principles and TB model calculations. We further demonstrated that, as the twist chirality is switched in the $p3$ structure, the signs of the Kerr angle for the AFM spin configuration as well as the polarization are inverted, which is also consistent with symmetry.

Our work highlights how the interplay between twisting and OC in AFM homo-bilayers represents an interesting ``knob'' for activating new physical properties not previously found in the untwisted case. 
Since our discussion is mainly based on symmetry arguments, a final comment on the possibility of experimentally detecting these effects is in order. Although the absolute values of the properties we studied in this work are rather small, this is not necessarily so in all analogous cases. 
Hence, further directions of study can be investigating analogous bilayers with different chemical compositions, where the effects can be enhanced in magnitude and the homogeneous AFM phase can be a true ground state. Our work demonstrated a general methodology for those analogous systems.

\section{Acknowledgement}
J.Y. acknowledges the support of the National Research Foundation of Korea (2020R1F1A1066548). 
Additional financial support in part by Samsung Electronics Co., Ltd. is also acknowledged.
K.Y. acknowledges the hospitality by CNR-SPIN c/o  Department of Physical and Chemical Science (DSFC) at University of L'Aquila (Italy) during the visit.

\begin{appendix}

\section{T1 and T2 twist} 
\label{sec:T1T2}
\textbf{T1 twist}: Starting with $p622$ layer-group symmetry, the 6-fold symmetry is lost by the decoration with OCs, because octahedra have no such symmetry, while the 3-fold axis is preserved, leaving three sites with 3-fold symmetry in all cases. The two possible low-symmetry layer groups are $p312$ and $p321$. If one starts with the RL stacking, T1 twisting yields three sites having stacking RL|LR|VV, the former two being connected by a 2-fold axis. We conclude that the symmetry is lowered to layer group $p321$. If one starts with LL stacking, T1 twisting results once again in three unique sites with exact 3-fold symmetry, with stacking LL|RR|VV. The 2-fold line through the 3-fold axes is preserved, resulting in layer-group symmetry $p312$.

\textbf{T2 twist}: The twisted honeycomb symmetry is $p321$. Starting with RL stacking, T2 twisting results in the three 3-fold sites having stacking RL|LV|VR. No 2-fold axis can exist, and the layer-group symmetry is $p3$. Note the special case of the untwisted bilayer, which has an additional inversion center and layer-group symmetry $p\bar{3}$. If one starts with LL stacking, T2 twisting results once again in three unique sites with exact 3-fold symmetry, with staking LL|RV|VR. The 2-fold axis in $p321$, connecting RV with VR is preserved, so the symmetry is not lowered in this case.

\section{Inequivalent Local Structural Properties Depending on OC Stacking in Presence of Twist: a geometric description} 
\label{sec:inequiv_stru}
\begin{figure}[t!]
\centering
\includegraphics[width=0.5\textwidth]{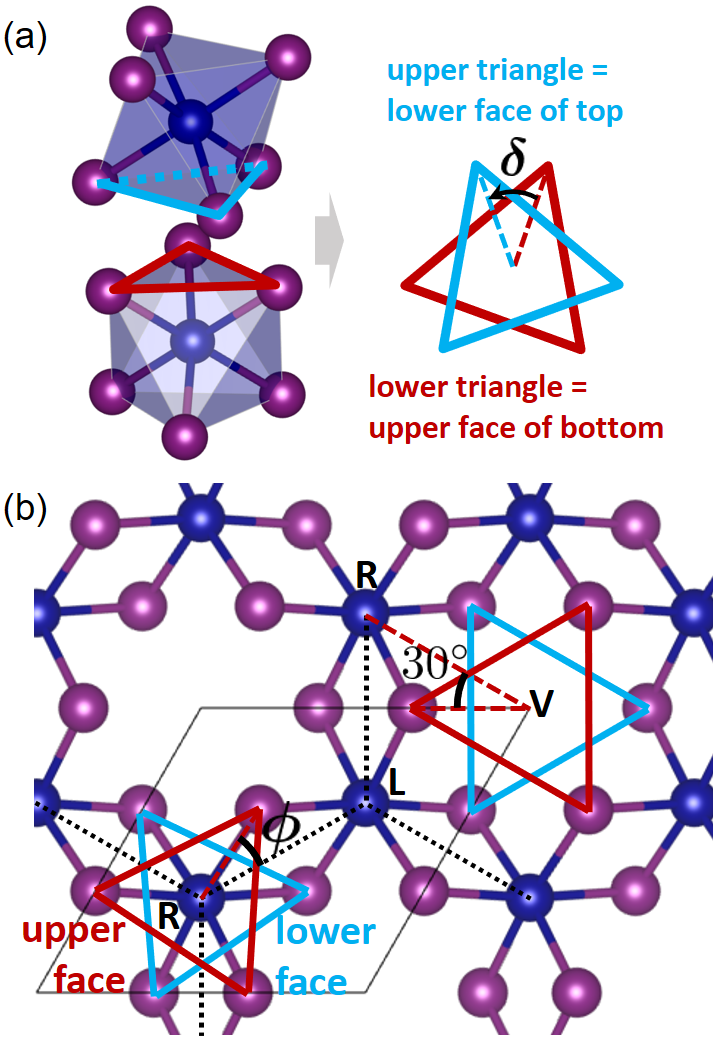}
\caption{
(a) Schematic picture representing the definition of $\delta$ in Appendix~\ref{sec:inequiv_stru}. (b) The upper and lower triangular faces of octahedra centered at Cr with R-OC (left two triangles) and V site (right) and their relative rotation angles with respect to the lattice are shown, where the black dotted line represents the honeycomb lattice. }
\label{fig:inequiv_env}
\end{figure}

Local structural properties at a given octahedra stacking can be represented by the bond lengths and angles between the I ions of the two adjacent triangular faces of the two octahedra. Without loss of generality, we will consider the ideal TBL CrI$_3$ systems without the structural relaxation (so-called rigid rotation).
We introduce an angle, $\delta$, as the relative counter-clockwise rotation angle of the lower face of the octahedron in the top layer (the upper one of the two adjacent triangles) with respect to the upper face of the bottom octahedron (the lower of the two), modulo 120$^\circ$ [See Fig.~\ref{fig:inequiv_env} (a)]. The different $\delta$ values determine the different local structural properties.
On the other hand, in a single octahedron, the upper face is rotated by $+\phi\approx 26.3^\circ$ and the lower face by $-\phi$ in the R octahedron, where the angle is defined in Fig.~\ref{fig:inequiv_env} (b). In the L octahedron, the sign is the opposite. 
For each OC stacking cases with the $+\theta$ twist, $\delta$ values are $2\phi + \theta$, $\theta$, $\theta$, and $-2\phi + \theta$ for LL, LR, RL, and RR, respectively. 
For a fixed OC at the top layer, for example, L, two different OCs at the bottom layer (L and R) give different local environments. Moreover, even the two cases of the same OC stacking, LL and RR, result in different local environments.  
One can find that the LR and RL have the same $\delta$ value. However, they are not identical when we consider the octahedra above and below. The different OCs give the different relative rotations of triangles within each octahedra, \textit{i.e.}, from top to bottom, $-2\phi\rightarrow\theta\rightarrow 2\phi$ in LR and $2\phi\rightarrow\theta\rightarrow-2\phi$ in RL. Nevertheless, RL and LR can be connected by the in-plane 2-fold rotation.

One can also consider the octahedron centered at the V site, which has larger triangular faces. In this case, the center-vertex line segment of the upper triangle deviates from the V-R line segment by $+30^\circ$, and the lower triangle is rotated from the upper one by $180^\circ$ [Fig.~\ref{fig:inequiv_env} (b)]. 
We can define $\delta$ in the same way for the V-OC stackings. Let us consider the possibilities appearing in the RL-T2 and LL-T2 TBLs. In the RL-T2 case, there exist LV stacking with $\delta = \phi +\theta +90^\circ$ and VR stacking with $\delta = -\phi +\theta -90^\circ$. Therefore, the local structural properties at these two stackings are not equivalent nor connected by any symmetry operation.
In the LL-T2 case, RV and VR stackings appear. For both RV and VR, $\delta = -\phi +\theta +90^\circ$, so that they can be connected by the in-plane 2-fold rotation. The $\delta$ values of VR in RL-T2 and LL-T2 are different due to the inverted OC configurations around the V site.

Let us apply this description to the OC stackings appearing in our specific cases in Fig.~\ref{fig:symmetry1}. In the LL-T1 case [Fig.~\ref{fig:symmetry1} (a)], the LL and RR stackings appear and their $\delta$ are -29.2$^\circ$, and -14.4$^\circ$, respectively. Therefore, two OC stackings are not equivalent. But the inequivalence does not reduce the symmetry in this case because the 2-fold rotation maps the OC stacking to itself. In the RL-T1 case [Fig.~\ref{fig:symmetry1} (b)], the RL and LR appear with $\delta=38.2^\circ$ for both, which means the two local properties are connected by 2-fold rotation. In the LL-T2 case [Fig.~\ref{fig:symmetry1} (c)], the stackings of OC and V site, the RV and VR can be found with $\delta=-34.5^\circ$ for both. Thus, their local properties of them are 2-fold symmetry-related. Finally, in the RL-T2 case [Fig.~\ref{fig:symmetry1} (d)], the LV with $\delta=18.1^\circ$ and VR with $\delta=25.5^\circ$ appear, implying inequivalent local environments.

\section{Practical Method to Generate TBL} 
\label{sec:cell_generating}
The twist angles which make the TBL systems commensurate are identified by $2\cos\theta = (m^2+n^2+4mn)/(m^2+n^2+mn)$ with a pair of positive coprime integers $(n,m)$~\cite{trambly_de_laissardiere_localization_2010}. Let us consider the case that the top layer is rotated by $\theta>0$ counter-clockwise, which we define as right-twist, and $n>m$. $21.79^\circ$ corresponds to $(n,m)=(2,1)$.
In practice, we use the alternative convention for the hexagonal cell that the angle $\gamma$ between the two in-plane lattice vectors is $60^\circ$ (in the main text, \textit{e.g.}, Fig.~\ref{fig:OCandTBL} (d), $\gamma=120^\circ$). Then, the primitive lattice vectors of the bottom layer are $\mathbf{a}^b_1 = a_0(1,0)$ and $\mathbf{a}^b_2 = a_0(1/2,\sqrt{3}/2)$. Those of the top layer are $\mathbf{a}^t_i = R(\theta)\mathbf{a}^b_i$ where $R(\theta)$ is rotation operator by $\theta$ around $z$-axis.
The lattice vectors of supercell, or the primitive cell of TBL, are $\mathbf{A}_1 = n\mathbf{a}_1^t + m\mathbf{a}_2^t = m\mathbf{a}_1^b + n\mathbf{a}_2^b$ and $\mathbf{A}_2 = -m\mathbf{a}_1^t + (n+m)\mathbf{a}_2^t = -n\mathbf{a}_1^b + (n+m)\mathbf{a}_2^b$. 
When the twist-axis passes through the  Cr or V site, 3-fold axes are at 0, 1/3, and 2/3 points of the long diagonal: integer multiples of
\begin{equation}
\begin{split}
\frac{1}{3}(\mathbf{A}_1+\mathbf{A}_2) &= \frac{1}{3}(n-m)\mathbf{a}_1^t + \frac{1}{3}(n+2m)\mathbf{a}_2^t\\
&= \frac{1}{3}(m-n)\mathbf{a}_1^b + \frac{1}{3}(2n+m)\mathbf{a}_2^b.
\end{split}
\end{equation}
If $n-m=3p$ where $p$ is a positive integer, the corresponding supercell is non-primitive, \textit{i.e.}, it represents a supercell of TBL cell.
For example, $38.21^\circ$ corresponds to $(n,m)=(4,1)$, however, the cell from this method is $\sqrt{3}\times\sqrt{3}$ supercell of primitive TBL cell.
If $n-m=3p+q$ where $q=1$ or $2$,
\begin{equation}
\label{eq:onethirdT}
\begin{split}
\frac{1}{3}(\mathbf{A}_1+\mathbf{A}_2) &= (p+\frac{1}{3}q)\mathbf{a}_1^t + (p+m+\frac{1}{3}q)\mathbf{a}_2^t\\
&= (-p-\frac{1}{3}q)\mathbf{a}_1^b + (2p+m+\frac{2}{3}q)\mathbf{a}_2^b.
\end{split}
\end{equation}
In this expression, the fractional number parts such as $\tfrac{1}{3}q$ determine the properties L, R, or V of the 3-fold axes sites of each layer according to the configurations in the primitive cell of each layer.

In the main text, the twist axis is at the Cr site of the AA-stacked bilayer. It can be represented by the case that one Cr atom is at the origin and the other is at $(1/3)(\mathbf{a}_1+\mathbf{a}_2)$. The LL bilayer can be obtained when the primitive cells of both the top and bottom layer have L-R-V at the 3-fold sites -- 0, 1/3, and 2/3 positions along the long diagonal. If the top layer's primitive cell is instead R-L-V, the bilayer becomes RL. 
However, to generate the T1 twist within a primitive TBL cell, we also consider the case that the twist axis is at the V site of the AA-stacked bilayer. When both layers are V-L-R, the bilayer is LL. If the top is V-R-L and the bottom is V-L-R, the bilayer is RL.

{\renewcommand{\arraystretch}{1.}
\begin{table}[b]
\begin{tabular}{C{0.04\textwidth}C{0.13\textwidth}C{0.02\textwidth}C{0.13\textwidth}m{0.07\textwidth}}
\hline
axis & \begin{tabular}{@{}c@{}}untwisted\\ stacking\end{tabular} & $q$ & \begin{tabular}{@{}c@{}}twisted\\ 3-fold sites\end{tabular} & type \\
\hline\hline
\multirow{4}{*}{Cr} & \multirow{2}{*}{LL|RR|VV} & 1 & LL|RV|VR & \multirow{2}{*}{T2 $p321$} \\
                    &                     & 2 & LL|VR|RV & \\
\cmidrule{2-5}
                    & \multirow{2}{*}{RL|LR|VV} & 1 & RL|LV|VR & \multirow{2}{*}{T2 $p3$} \\
                    &                     & 2 & RL|VR|LV & \\
\hline
\multirow{4}{*}{V } & \multirow{2}{*}{VV|LL|RR} & 1 & VV|LR|RL & \multirow{2}{*}{T1 $p321$} \\
                    &                     & 2 & VV|RL|LR & \\
\cmidrule{2-5}
                    & \multirow{2}{*}{VV|RL|LR} & 1 & VV|RR|LL & \multirow{2}{*}{T1 $p312$} \\
                    &                     & 2 & VV|LL|RR & \\
\hline
\end{tabular}
\caption{
The 3-fold sites configurations (top and bottom layer's labels at 0, 1/3, and 2/3 of the long diagonal), twist series, and symmetries of the TBL according to the cell generation settings. One can also derive the settings for the bare honeycomb cases without OCs, $p622$ (T1) or $p321$ (T2), from the table.}
\label{tab:appdx_symmetry}
\end{table}}

Let us take the LL bilayer with the twist axis at the Cr site as an example. When the twist angle is given by $q=1$ cases, Eq.~(\ref{eq:onethirdT}) indicates that 1/3 point of the TBL cell's long diagonal corresponds to 1/3 (2/3) point of the primitive cell's long diagonal of the top (bottom) layer. In our settings, it is R (V) for the top (bottom) layer. Similarly, 2/3 point of TBL's long diagonal corresponds to 2/3 for the top (V) and 1/3 for the bottom (R), respectively. We denote it as LL|RV|VR according to the convention introduced in Sec.~\ref{sec:str_sym}. This is the T2 twist with $p321$ symmetry. Likewise, one can examine the 3-fold sites along the long diagonal for other stacking and twist cases and determine the symmetry of TBLs. The results are summarized in Table~\ref{tab:appdx_symmetry}.
When $q=0$, $(1/3)(\mathbf{A}_1+\mathbf{A}_2)$ becomes a new lattice vector having the $\mathbf{A}_1$ as a new long diagonal. In this case, the TBL is a T1 twist even for the Cr site twisting. On the other hand, V site twisting becomes T1 for every $q$.

The opposite twist chirality case can also be analyzed in the same way but using $-\theta$ and exchanged $n$ and $m$ values ($m-n=3p+q$). Using the same settings for RL bilayer with Cr site twisting, the resulting TBL structures are RL|VR|LV for $q=1$ and RL|LV|VR for $q=2$. The mirror image relation can be confirmed from this derivation.
It also provides a practical way to produce the primitive TBL cell of T1 twist from the V site twisting. The $38.21^\circ$ twist is equivalent to $(38.21+120)^\circ = (180-21.79)^\circ$ twist, and $180^\circ$ twist exchanges two sublattices. As a consequence, $-21.79^\circ$ twist of RL (LL) bilayer around V site is equivalent to the $38.21^\circ$ twist of LL (RL) around Cr site after appropriate origin shift following the twist. Such structures are the primitive TBL cell from $(n,m)=(1,2)$. 

\end{appendix}

\bibliography{TBL_CrI3}

\begin{thebibliography}{87}
\bibitem [{foo()}]{footnote1_OC}%
  \BibitemOpen
  \href@noop {} {}\bibinfo {note} {In inorganic chemistry, these two
  configurations are generally referred to as $\Lambda$ and $\Delta$ isomers
  for L and R, respectively, forming two enantiomers of an octahedral complex;
  See also Ref.~\cite{NomenclatureInorgChem_Ch9}.}\BibitemShut {Stop}%
\bibitem [{b_f()}]{b_footnote2_3foldsites}%
  \BibitemOpen
  \href@noop {} {}\bibinfo {note} {These 3-fold sites orders are permutable by
  rotating or translating the defined unitcell. In other words, a permutation
  of the notation still represents the same system. Moreover, the set of all
  the possibilities of this notation exhausts all the possible symmetries with
  3-fold rotation.}\BibitemShut {Stop}%
\bibitem [{d_f()}]{d_footnote4_hBN}%
  \BibitemOpen
  \href@noop {} {}\bibinfo {note} {In the hBN case, such local polar structures
  appear only in the small-twist angle case. In the CrI$_3$ case, the local
  stacking structures in a small-twist angle are still non-polar. However, the
  polarization mechanism in our work is still valid for a large-twist angle.
  The hBN can be polar at a large-twist angle by an analogous global symmetry
  lowering, but without a switchability. The difference originates from the
  different behaviors of the sublattice degree of freedoms in each case under
  the improper rotations.}\BibitemShut {Stop}%
\bibitem [{c_f()}]{c_footnote3_inv}%
  \BibitemOpen
  \href@noop {} {}\bibinfo {note} {One can construct the enantiomeric pair by
  adopting an inversion operator instead of $\mathcal{M}_z$, with an additional
  180$^\circ$ rotation. It does not alter our key findings. Similarly, any
  improper rotation which exchanges the two layers can be used. In the case of
  a vertical mirror, which is an improper rotation but not layer-exchanging,
  the mirror image is a $-\theta$ twisted system but with the opposite OC
  stacking.}\BibitemShut {Stop}%
\end{thebibliography}

\clearpage

\onecolumngrid
\vspace{\columnsep}
\begin{center}
\textbf{\large Supplemental material for ``Activating magnetoelectric optical properties by twisting antiferromagnetic bilayers''}
\end{center}

\vspace{\columnsep}
\twocolumngrid

\setcounter{equation}{0}
\setcounter{figure}{0}
\setcounter{table}{0}
\setcounter{page}{1}
\setcounter{section}{0}
\makeatletter

\renewcommand{\thetable}{S\Roman{table}}
\renewcommand{\thefigure}{S\arabic{figure}}
\renewcommand{\thesection}{S\Roman{section}}
\renewcommand{\theequation}{S\arabic{equation}}

\section{Details of the Tight Binding Model}
\subsection{Structure and Basis for the Monolayer}
\begin{figure}[t!]
\centering
\includegraphics[width=0.5\textwidth]{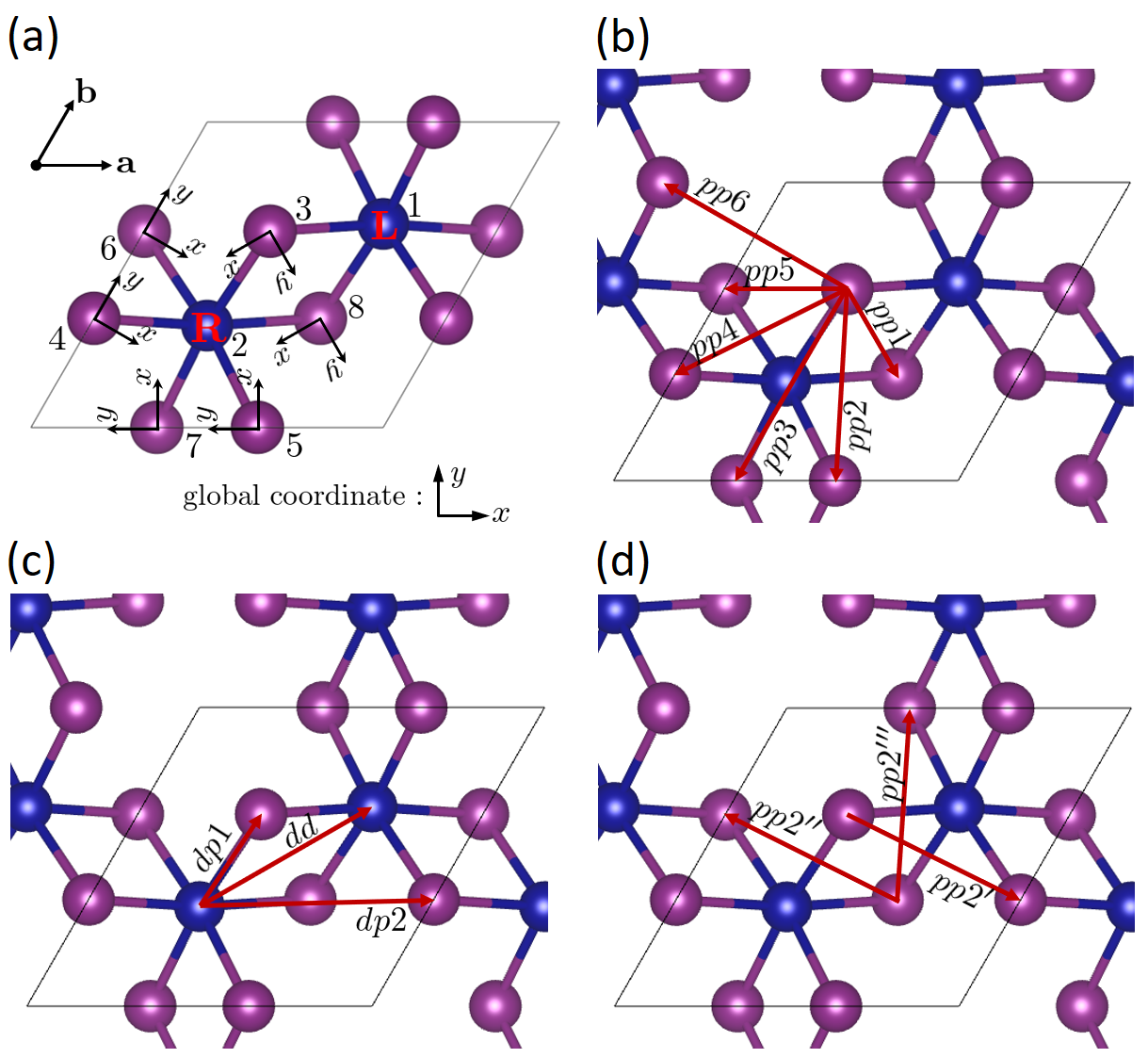}
\caption{
Structure of the monolayer CrI$_3$. Atom numbers, OCs for each Cr, and local coordinates for each I are shown in (a).
Selected hoppings between the atoms for the TB model are shown in (b-d). (b) I-I hoppings. (c) Cr-I and Cr-Cr hoppings. (d) Examples of I-I hoppings symmetrically related to the hoppings in (b).}
\label{fig:basis_hoppings}
\end{figure}

First, we construct the TB Hamiltonian for the monolayer and then extend it to the bilayer or TBLs. For consistency with the TBL description in Appendix C of the main text, the angle $\gamma$ between the lattice vectors $\mathbf{a}$ and $\mathbf{b}$ is $60^\circ$ in this Supplemental Material. Therefore, the structure used for the TB model and the DFT calculation for generating the Wannier Hamiltonian is defined using the following lattice vectors and the fractional coordinates listed in Table~\ref{tab:fractional_coord}.
\begin{equation}
\begin{split}
\mathbf{a} &= a(1,0,0) \\
\mathbf{b} &= a(1/2,\sqrt{3}/2,0) \\
\mathbf{c} &= c(0, 0, 1) \\
\end{split}
\end{equation}
where $a = 6.98525$ \AA\ and $c = 24$ \AA\ including a vacuum.
The structure is shown in Fig.~\ref{fig:basis_hoppings} (a).
Here, Cr1 corresponds to the L-OC and Cr2 to the R-OC.

\begin{table}
\begin{tabular}{C{0.02\textwidth}C{0.05\textwidth}C{0.12\textwidth}C{0.12\textwidth}C{0.12\textwidth}}
\hline
 No. & spc. & $\mathbf{a}$ & $\mathbf{b}$ & $\mathbf{c}$  \\
\hline
 1 &  Cr & 0.6666666667 &   0.6666666667 &   0.500000000   \\
 2 &  Cr & 0.3333333333 &   0.3333333333 &   0.500000000   \\
 3 &  I  & 0.3572400000 &   0.6427600000 &   0.565970000   \\
 4 &  I  & 0.0000000000 &   0.3572400000 &   0.565970000   \\
 5 &  I  & 0.6427600000 &   0.0000000000 &   0.565970000   \\
 6 &  I  & 0.0000000000 &   0.6427600000 &   0.434030000   \\
 7 &  I  & 0.3572400000 &   0.0000000000 &   0.434030000   \\
 8 &  I  & 0.6427600000 &   0.3572400000 &   0.434030000   \\
\hline
\end{tabular}
\caption{
Atom numbers, species, and their fractional coordinates with respect to the ($\mathbf{a}$,$\mathbf{b}$,$\mathbf{c}$) in a monolayer unit cell.}
\label{tab:fractional_coord}
\end{table}

For the Cr, $d$-orbitals defined in the global reference system are chosen as a basis for the TB model. For I atoms, $p$-orbitals are defined in the local coordinates as shown in Fig.~\ref{fig:basis_hoppings}. 
The local coordinates are rotated by $-150^\circ$ for I3 and I8, $-30^\circ$ for I4 and I6, and $90^\circ$ for I5 and I7. The orbitals will be further rotated in the TBL systems, in which the $d$-orbitals will no longer be aligned to the global reference system.

Apart from the radial factor, $d$-orbitals are described by the following defining equations:
\begin{equation}
\label{eq:dorbitals}
\begin{split}
&\ket{d_{z^2}} \sim \tfrac{1}{\sqrt{6}}(2z^2-x^2-y^2) \\
&\ket{d_{x^2-y^2}} \sim \tfrac{1}{\sqrt{2}}(x^2-y^2) \\
&\ket{d_{xy}} \sim \sqrt{2}xy \\
&\ket{d_{xz}} \sim \sqrt{2}xz \\
&\ket{d_{yz}} \sim \sqrt{2}yz \\
\end{split}
\end{equation}
Symmetry properties can be easily derived from these expressions.
In this work, matrices are defined with respect to the $d$-orbitals basis set in the order given above. For $p$-orbitals, $p_i \sim x_i$ in each local coordinates with basis order $\{p_x,p_y,p_z\}$.

\subsection{DFT and Wannierization}
In order to define the Wannier Hamiltonian [S1], we use the same DFT methods as the main text, except that the $9\times9\times1$ $k$-point grid is used for the monolayer and untwisted bilayer. The Wannier $k$-grid is also $9\times9\times1$.
The basis for the TB model is also used as initial projectors for the maximally localized Wannier functions. In order to include the spin degrees of freedoms and SOC, we use the spinor Wannier functions. 
In our calculation, the Wannier interpolation method reproduces the DFT band structure shown in Fig~\ref{fig:bands_mono} very well. Thus, the Wannier and DFT bands are used interchangeably hereafter.
The resulting Wannier functions, after spread minimization, maintain the characteristics of the orbitals defining the basis.

\subsection{Tight Binding Formalism}
In the TB theory, Hamiltonian is described by the hopping parameters from one orbital to another. In general, $\mel{\phi^{\sigma}_{i\alpha}(\mathbf{r}-\boldsymbol{\tau}_{i})}{\hat{H}}{\phi^{\sigma'}_{j\beta}(\mathbf{r}-\mathbf{R}-\boldsymbol{\tau}_{j})}$
represents the hopping from the orbital $\alpha$ with spin $\sigma$ of the atom $i$ located at $\boldsymbol{\tau}_{i}$ to the orbital $\beta$ with spin $\sigma'$ of the atom $j$ at $\mathbf{R}+\boldsymbol{\tau}_{j}$, where $\mathbf{R}$ is a lattice vector.
The index of the basis set of the TB Hamiltonian is represented as the combined index $(\sigma i \alpha)$ of the spin, atom, and orbital. 
The matrix element of the Hamiltonian at a specific $\mathbf{k}$ is given as 
\begin{equation}
\begin{split}
&(H_{\mathbf{k}})_{(\sigma i\alpha)(\sigma' j\beta)} = \\
&\sum_{\mathbf{R}}e^{i\mathbf{k}\cdot(\mathbf{R}+\boldsymbol{\tau}_{j}-\boldsymbol{\tau}_{i})}\mel{\phi^{\sigma}_{i\alpha}(\mathbf{r}-\boldsymbol{\tau}_{i})}{\hat{H}}{\phi^{\sigma'}_{j\beta}(\mathbf{r}-\mathbf{R}-\boldsymbol{\tau}_{j})}.
\end{split}
\end{equation}
The following equation determines the energy eigenvalues $E_{\mathbf{k}n}$ and eigenvectors $\mathbf{C}_{\mathbf{k}n}$ where $n$ is the band index.
\begin{equation}
H_{\mathbf{k}}\mathbf{C}_{\mathbf{k}n} = E_{\mathbf{k}n}\mathbf{C}_{\mathbf{k}n}    
\end{equation}
In this formalism, $\mathbf{C}_{\mathbf{k}n}$ corresponds to cell-periodic part of the Bloch wave-function, $u_{\mathbf{k}n}(\mathbf{r})$, while the full Bloch wave-function is $\psi_{\mathbf{k}n}(\mathbf{r})=e^{i\mathbf{k}\cdot\mathbf{r}}u_{\mathbf{k}n}(\mathbf{r})$ [S2].
The velocity operator is replaced with the effective tight-binding velocity operator,  
$(\hat{\mathbf{v}}^{\mathbf{k}})^{\text{TB}} = (1/\hbar)\boldsymbol{\nabla}_{\mathbf{k}}H_{\mathbf{k}}$ [S3].
Its matrix element is
\begin{equation}
\begin{split}
(\hat{\mathbf{v}}^{\mathbf{k}})^{\text{TB}}_{(\sigma i\alpha)(\sigma' j\beta)} &=\frac{1}{\hbar}
\sum_{\mathbf{R}}e^{i\mathbf{k}\cdot(\mathbf{R}+\boldsymbol{\tau}_{j}-\boldsymbol{\tau}_{i})}i(\mathbf{R}+\boldsymbol{\tau}_{j}-\boldsymbol{\tau}_{i}) \\
&\times\mel{\phi^{\sigma}_{i\alpha}(\mathbf{r}-\boldsymbol{\tau}_{i})}{\hat{H}}{\phi^{\sigma'}_{j\beta}(\mathbf{r}-\mathbf{R}-\boldsymbol{\tau}_{j})}.
\end{split}
\end{equation}
On the other hand, when we calculate the MOKE spectrum from the DFT via Wannier functions, the velocity operator in the commutator form $\hat{\mathbf{v}}=(i/\hbar)[\hat{H},\hat{\mathbf{x}}]$ is used. The Hamiltonian and the position matrices in the Wannier function basis are used to evaluate this quantity.

\subsection{Monolayer Hamiltonian}
Here we construct the Hamiltonian of the monolayer CrI$_3$ in the TB theory framework as described in the previous subsection.
In practice, we truncate the hoppings so that only those between the nearby atoms are included. The allowed hoppings are shown in Fig.~\ref{fig:basis_hoppings} (b-d).

Let us first consider the intra-atomic Hamiltonian parameters. Single orbital energy levels are $\epsilon^{\sigma}_{i\alpha}=\mel{\phi^{\sigma}_{i\alpha}}{\hat{H}}{\phi^{\sigma}_{i\alpha}}$. In the potential affected by the presence of other atoms, $\epsilon^{\sigma}_{i\alpha}$ can differ for each orbital. For the $d$-orbitals, 3-fold symmetry imposes the conditions $\epsilon^{\sigma}_{x^2-y^2}=\epsilon^{\sigma}_{xy}$ and $\epsilon^{\sigma}_{xz}=\epsilon^{\sigma}_{yz}$.
In addition, the absence of $\mathcal{M}_z$ symmetry and the presence of $D_3$ point-group symmetry allow inter-orbital hopping terms $\mel{d^{\sigma}_{x^2-y^2}}{\hat{H}}{d^{\sigma}_{xz}} = -\mel{d^{\sigma}_{xy}}{\hat{H}}{d^{\sigma}_{yz}} = \delta^{\sigma}_d$.
In summary, the intra-atomic Hamiltonian for $d$-orbitals within one spin channel is expressed as the following matrix:
\begin{equation}
h_{d} = 
\begin{pmatrix}
 \varepsilon_{z^2} & 0  & 0  & 0  & 0  \\
 0  & \varepsilon_{xy} & 0  & \delta_d  & 0  \\
 0  & 0  & \varepsilon_{xy} & 0  & -\delta_d  \\
 0  & \delta_d  & 0  & \varepsilon_{xz} & 0  \\
 0  & 0  & -\delta_d  & 0  & \varepsilon_{xz} \\
\end{pmatrix}
\end{equation}
The obtained parameters are the followings in eV unit,
\begin{equation}
\begin{split}
&\varepsilon^{\uparrow}_{z^2} = 2.3150\\
&\varepsilon^{\uparrow}_{xy} = 2.1789\\
&\varepsilon^{\uparrow}_{xz} = 1.4819\\
&\delta^{\uparrow}_{d} = 0.3414\\
\end{split}
\end{equation}
for spin-up, and 
\begin{equation}
\begin{split}
&\varepsilon^{\downarrow}_{z^2} = -2.7172\\
&\varepsilon^{\downarrow}_{xy} = -2.2970\\
&\varepsilon^{\downarrow}_{xz} = -0.7964\\
&\delta^{\downarrow}_{d} = -0.8568\\
\end{split}
\end{equation}
for spin-down.
Similarly, for $p$-orbitals,
\begin{equation}
h_{p} = 
\begin{pmatrix}
 \varepsilon_{x} & 0  & 0 \\
 0  & \varepsilon_{y} & \delta_p \\
 0  & \delta_p & \varepsilon_{z} \\
\end{pmatrix}
\end{equation}
where the forbidden terms are due to the mirror symmetry.
The parameters are 
\begin{equation}
\begin{split}
&\varepsilon^{\uparrow}_{x} = -2.2879\\
&\varepsilon^{\uparrow}_{y} = -1.7891\\
&\varepsilon^{\uparrow}_{z} = -1.9341\\
&\delta^{\uparrow}_p = 0.2754\\
\end{split}
\end{equation}
for spin-up, and 
\begin{equation}
\begin{split}
&\varepsilon^{\downarrow}_{x} = -2.3158\\
&\varepsilon^{\downarrow}_{y} = -1.8468\\
&\varepsilon^{\downarrow}_{z} = -1.9866\\
&\delta^{\downarrow}_p = 0.2531\\
\end{split}
\end{equation}
for spin-down.
The single orbital energy levels are expressed with respect to the Fermi level.

The SOC is considered as $\lambda\mathbf{L}\cdot\mathbf{S}$ perturbation terms within a single atom for both Cr and I.
SOC is the only interaction that connects the spin-up and spin-down orbitals.
Spin operators are
\begin{equation}
\label{eq:sx}
S_{x} = \frac{\hbar}{2} 
\begin{pmatrix}
 0 & 1 \\
 1 & 0 \\
\end{pmatrix}
\end{equation}
\begin{equation}
S_{y} = \frac{\hbar}{2} 
\begin{pmatrix}
 0 & -i \\
 i & 0 \\
\end{pmatrix}
\end{equation}
\begin{equation}
S_{z} = \frac{\hbar}{2} 
\begin{pmatrix}
 1 & 0 \\
 0 & -1 \\
\end{pmatrix}.
\end{equation}
Orbital angular momentum operators are 
\begin{equation}
L^{p}_{x} = \hbar 
\begin{pmatrix}
 0 & 0  & 0 \\
 0 & 0  &-i \\
 0 & i  & 0 \\
\end{pmatrix}
\end{equation}
\begin{equation}
L^{p}_{y} = \hbar
\begin{pmatrix}
 0 & 0  & i \\
 0 & 0  & 0 \\
-i & 0  & 0 \\
\end{pmatrix}
\end{equation}
\begin{equation}
L^{p}_{z} = \hbar
\begin{pmatrix}
 0 & -i & 0 \\
 i & 0  & 0 \\
 0 & 0  & 0 \\
\end{pmatrix}
\end{equation}
for $p$-orbitals and
\begin{equation}
L^{d}_{x} = \hbar 
\begin{pmatrix}
 0  & 0  & 0  & 0  & i\sqrt{3} \\
 0  & 0  & 0  & 0  & i         \\
 0  & 0  & 0  & -i & 0         \\
 0  & 0  & i  & 0  & 0         \\
-i\sqrt{3} & -i & 0  & 0  & 0         \\
\end{pmatrix}
\end{equation}
\begin{equation}
L^{d}_{y} = \hbar 
\begin{pmatrix}
 0        & 0  & 0  &-i\sqrt{3}& 0 \\
 0        & 0  & 0  & i        & 0 \\
 0        & 0  & 0  & 0        & i \\
 i\sqrt{3}&-i  & 0  & 0        & 0 \\
 0        & 0  &-i  & 0        & 0 \\
\end{pmatrix}
\end{equation}
\begin{equation}
\label{eq:ldz}
L^{d}_{z} = \hbar 
\begin{pmatrix}
 0  & 0  & 0  & 0  & 0  \\
 0  & 0  & -2i& 0  & 0  \\
 0  & 2i & 0  & 0  & 0  \\
 0  & 0  & 0  & 0  &-i  \\
 0  & 0  & 0  & i  & 0  \\
\end{pmatrix}
\end{equation}
for $d$-orbitals.
SOC parameters are $\hbar^2\lambda_p = 0.6242$ eV for I and $\hbar^2\lambda_d = 0.0507$ eV for Cr.

Since we adopt the orbitals defined in the rotated local coordinates and the twist further rotates the coordinates, one should transform the basis of the SOC Hamiltonian from the orbitals defined in the global coordinates to those in the local coordinates, \textit{i.e.}, passive transformation. 
\begin{equation}
\begin{split}
&\mel{\phi^{\sigma}_{\alpha}}{\hat{H}^{\text{SOC}}}{\phi^{\sigma'}_{\beta}} \\
&= \sum_{\gamma'\delta'}\braket{\phi^{\sigma}_{\alpha}}{\phi^{\sigma}_{\gamma'}}\mel{\phi^{\sigma}_{\gamma'}}{\hat{H}^{\text{SOC}}}{\phi^{\sigma'}_{\delta'}}\braket{\phi^{\sigma'}_{\delta'}}{\phi^{\sigma'}_{\beta}}
\end{split}
\end{equation}
where the unprimed $\alpha$ and $\beta$ are the orbital indices in the global coordinates and the primed $\gamma'$ and $\delta'$ are those in the local coordinates. The identity relation of the orbitals $I=\sum_{\alpha}\ket{\phi_{\alpha}}\bra{\phi_{\alpha}}$, where the summation is over the same orbital species ($d$ or $p$), is used.
It can be written in the matrix form, $H^{\text{SOC}}_{\sigma\sigma'} = R H'^{\text{SOC}}_{\sigma\sigma'} R^{T}$, where $R_{\alpha\gamma'} \equiv \braket{\phi_{\alpha}}{\phi_{\gamma'}}$. $H^{\text{SOC}}$ is the SOC matrix in the global coordinates which is constructed by considering Eq.~(\ref{eq:sx}-\ref{eq:ldz}) as the operators in the global coordinates.
Thus, the SOC matrix in the local coordinates is $H'^{\text{SOC}}_{\sigma\sigma'} = R^{T} H^{\text{SOC}}_{\sigma\sigma'} R$. This transformation is applied for each $\uparrow\uparrow$, $\uparrow\downarrow$, $\downarrow\uparrow$, and $\downarrow\downarrow$ spin blocks. 
For $p$-orbitals, $R$ matrix is easily obtained from the relation $\ket{p_{\alpha}}\propto x_{\alpha}$.
\begin{equation}
\label{eq:Rp}
 R_{p}(\theta)=
\begin{pmatrix}
 \cos\theta & -\sin\theta  & 0 \\
 \sin\theta &  \cos\theta  & 0 \\
 0 & 0  & 1 \\
\end{pmatrix}
\end{equation}
where $\theta$ is the rotation angle of the local coordinates with respect to the global coordinates.
The transformation for $d$-orbitals is required when we investigate the twisted bilayers. 
The following relations between the rotated and unrotated $d$-orbitals are obtained from Eq.~(\ref{eq:dorbitals}).
\begin{equation}
\begin{split}
&\ket{d_{z'^2}} = \ket{d_{z^2}} \\
&\ket{d_{x'^2-y'^2}} = \cos2\theta\ket{d_{x^2-y^2}} + \sin2\theta\ket{d_{xy}}\\
&\ket{d_{x'y'}} = -\sin2\theta\ket{d_{x^2-y^2}} + \cos2\theta\ket{d_{xy}} \\
&\ket{d_{x'z'}} = \cos\theta\ket{d_{xz}} + \sin\theta\ket{d_{yz}} \\
&\ket{d_{y'z'}} = -\sin\theta\ket{d_{xz}} + \cos\theta\ket{d_{yz}}  \\
\end{split}
\end{equation}
Therefore,
\begin{equation}
R_{d}(\theta)=
\begin{pmatrix}
1 & 0 & 0 & 0 & 0 \\
0 & \cos2\theta & -\sin2\theta & 0 & 0 \\
0 & \sin2\theta &  \cos2\theta & 0 & 0 \\
0 & 0 & 0 & \cos\theta & -\sin\theta \\
0 & 0 & 0 & \sin\theta &  \cos\theta \\
\end{pmatrix}.
\end{equation}

Inter-atomic hoppings are constructed for each type of hoppings. First, hoppings between $p$-orbitals on I's are directly extracted from Wannier interpolated Hamiltonian, being averaged among the equivalent hoppings and symmetrized.
Since we adopted local coordinates for $p$-orbitals, parameters obtained for one hopping path can be used for the symmetry-related hopping paths. The selected $p$-$p$ hoppings for the model are shown in Fig.~\ref{fig:basis_hoppings} (b), or symmetry-related to them.

The obtained parameters are as below in the eV unit. The parameters  $\mel{p^{\sigma}_{i\alpha}}{\hat{H}}{p^{\sigma}_{j\beta}}$ are written as matrices for each $(i,j)$ pair which corresponds to one of $pp1$-$pp6$ paths and has $(\alpha,\beta)$ as the index of matrix elements.
For spin-up,
\begin{equation}
h^{\uparrow}_{pp1} = 
\begin{pmatrix}
 -0.1883  & 0        &  0 \\
 0        & 0.2276   & -0.4329 \\
 0        & -0.4329  &  0.4961 \\
\end{pmatrix}
\end{equation}
\begin{equation}
h^{\uparrow}_{pp2} = 
\begin{pmatrix}
-0.4026  & -0.0620  &  0.0444 \\
-0.4922  & 0.0325   &  0.0474 \\
0.0014   & 0.0174   & -0.0888 \\
\end{pmatrix}
\end{equation}
\begin{equation}
h^{\uparrow}_{pp3} = 
\begin{pmatrix}
 0.0351  & -0.0426  & -0.0083 \\
 0.0426  & 0.0485   & -0.0048 \\
 0.0083  & -0.0048  & -0.0662 \\
\end{pmatrix}
\end{equation}
\begin{equation}
h^{\uparrow}_{pp4} = (h^{\uparrow}_{pp2})^{T}
\end{equation}
\begin{equation}
h^{\uparrow}_{pp5} = 
\begin{pmatrix}
 -0.2064  & -0.0225 & -0.2649 \\
 0.0225   & 0.0440  & 0.0802  \\
 0.2649   & 0.0802  & 0.3056  \\
\end{pmatrix}
\end{equation}
\begin{equation}
h^{\uparrow}_{pp6} = 
\begin{pmatrix}
 0.0581  & 0.1056   & -0.0091 \\
 -0.1056 & -0.2437  & -0.0095 \\
 0.0091  & -0.0095  & -0.0375 \\
\end{pmatrix}.
\end{equation}
For spin-down,
\begin{equation}
h^{\downarrow}_{pp1} = 
\begin{pmatrix}
 -0.2465  & 0        & 0 \\
 0        & 0.2159   & -0.4749 \\
 0        & -0.4749  & 0.5873 \\
\end{pmatrix}
\end{equation}
\begin{equation}
h^{\downarrow}_{pp2} = 
\begin{pmatrix}
 -0.4279 & -0.0662 & 0.0587 \\
 -0.5198 & 0.0427  & 0.0404 \\
 0.0156  & 0.0268  & -0.1057 \\
\end{pmatrix}
\end{equation}
\begin{equation}
h^{\downarrow}_{pp3} = 
\begin{pmatrix}
 0.0236  & -0.0243  & -0.0006 \\
 0.0243  & 0.0198   & -0.0013 \\
 0.0006  & -0.0013  & -0.0343 \\
\end{pmatrix}
\end{equation}
\begin{equation}
h^{\downarrow}_{pp4} = (h^{\downarrow}_{pp2})^{T}
\end{equation}
\begin{equation}
h^{\downarrow}_{pp5} = 
\begin{pmatrix}
 -0.2315  & -0.0338  & -0.2829 \\
 0.0338   & 0.0759   & 0.1032 \\
 0.2829   & 0.1032   & 0.3282 \\
\end{pmatrix}
\end{equation}
\begin{equation}
h^{\downarrow}_{pp6} = 
\begin{pmatrix}
0.0475  & 0.1017   & -0.0107 \\
-0.1017 & -0.2363  & -0.0021 \\
0.0107  & -0.0021  & -0.0291 \\
\end{pmatrix}.
\end{equation}

Other I-I hoppings are related by symmetry to these hoppings, as shown by some examples in Fig.~\ref{fig:basis_hoppings} (d).
The path $ppN'$ ($N=1,...,6$) is the mirror-symmetric pair of $ppN$, where the mirror plane is the vertical plane including $pp1$ hopping path. By mirror symmetry and local coordinates, the relation between $ppN'$ and $ppN$ is given as follows: 
\begin{equation}
h_{ppN'} = 
\begin{pmatrix}
-1 & 0 & 0 \\
0 & 1 & 0 \\
0 & 0 & 1 \\
\end{pmatrix}
h_{ppN}
\begin{pmatrix}
-1 & 0 & 0 \\
0 & 1 & 0 \\
0 & 0 & 1 \\
\end{pmatrix}.
\end{equation}
Tha path $ppN''$ is the 2-fold rotation pair of $ppN$, where the rotation axis is a line connecting Cr1 and Cr2. Similarly,
\begin{equation}
h_{ppN''} = 
\begin{pmatrix}
1 & 0 & 0 \\
0 & -1 & 0 \\
0 & 0 & -1 \\
\end{pmatrix}
h_{ppN}
\begin{pmatrix}
1 & 0 & 0 \\
0 & -1 & 0 \\
0 & 0 & -1 \\
\end{pmatrix}.
\end{equation}
The path $ppN'''$ is the mirror symmetric pair of $ppN''$. Combining former two cases, 
\begin{equation}
h_{ppN'''} = h_{ppN}.
\end{equation}
The remaining I-I hoppings are related to these by 3-fold rotation, which does not transform the hopping parameter matrices.

The $d$-$p$ hopping and $d$-$d$ hopping are expressed in terms of the Slater-Koster parametrization with the directional cosines $l$, $m$, and $n$. One can find the details of the parametrization in their original article, Ref. [S4].
Slater-Koster parameters should be rotated properly for the $p$-orbitals in the local coordinates.
Let us consider the hopping from the orbital $\phi_{i\alpha}$ defined in the coordinate $(\hat{\mathbf{x}},\hat{\mathbf{y}},\hat{\mathbf{z}})$ to the $p$-orbital $p_{j\beta'}$ defined in the coordinate $(\hat{\mathbf{x}}',\hat{\mathbf{y}}',\hat{\mathbf{z}}')$, $\mel{\phi_{i\alpha}}{\hat{H}}{p_{j\beta'}}$.
In a similar way as the SOC matrix case, $\mel{\phi_{i\alpha}}{\hat{H}}{p_{j\beta'}} = \sum_{\gamma}\mel{\phi_{i\alpha}}{\hat{H}}{p_{j\gamma}}\braket{p_{j\gamma}}{p_{j\beta'}}$ and $R_{\gamma\beta'} = \braket{p_{\gamma}}{p_{\beta'}} = \hat{\mathbf{x}}_{\gamma}\cdot\hat{\mathbf{x}}'_{\beta}$.
Then we can express the rotation of the hopping parameter matrix as $h'=hR$ where the $h'$ is the matrix defined with the local coordinates, and the $h$ is the matrix defined by the Slater-Koster parametrization in a single global coordinate.

\begin{figure}[t!]
\centering
\includegraphics[width=0.5\textwidth]{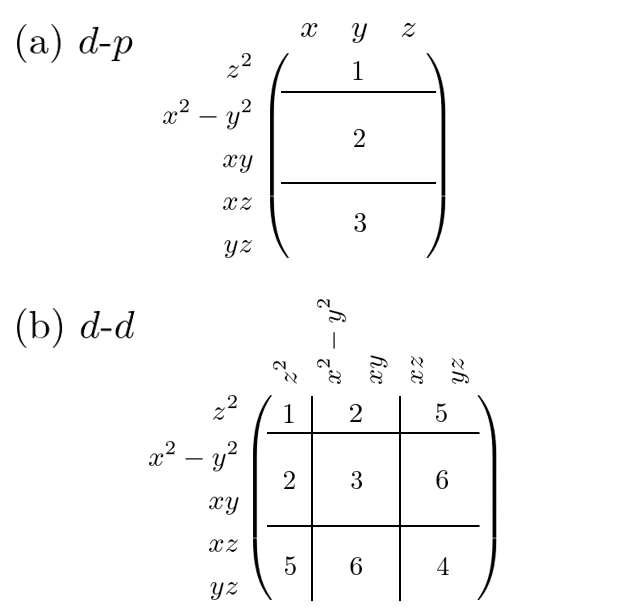}
\caption{
Division of the blocks of (a) $d$-$p$ and (b) $d$-$d$ hopping matrix for the blockwise Slater-Koster parameters.}
\label{fig:blocks}
\end{figure}

In order to improve the fitting, the Slater-Koster parameters are determined blockwisely. The division of the blocks is depicted in Fig~\ref{fig:blocks}. The division is determined by the same consideration for symmetry as the $d$-orbital energy levels of a single Cr ion.
Let us take the $d$-$p$ hopping as an example.
First, using $V^{(1)}_{dp\sigma}$ and $V^{(1)}_{dp\pi}$ which are the Slater-Koster parameters for the block 1, construct $h^{(1)}$ matrix in the global coordinate.
Next, rotate it to obtain the matrix in the local coordinate, \textit{i.e.}, $h'^{(1)}=h^{(1)}R$.
From this $h'^{(1)}$ matrix, get $\mel{d_{z^2}}{\hat{H}}{p_{x'_i}}$ elements of $h'$.
Repeat this procedure for $d_{x^2-y^2}$ and $d_{xy}$ with $V^{(2)}_{dp\sigma}$ and $V^{(2)}_{dp\pi}$, and for $d_{xz}$ and $d_{yz}$ with $V^{(3)}_{dp\sigma}$ and $V^{(3)}_{dp\pi}$.
Determination of those parameters is the inverse process of it.
The followings are resultant blockwise parameters for $d$-$p$ hoppings ($dp1$ and $dp2$) in eV.
For spin-up,
\begin{equation}
\begin{split}
&V^{(1)\uparrow}_{dp1\sigma} = -14.4721 \\
&V^{(1)\uparrow}_{dp1\pi} = 0.6701 \\
&V^{(2)\uparrow}_{dp1\sigma} = -0.7517 \\
&V^{(2)\uparrow}_{dp1\pi} = 0.7405 \\
&V^{(3)\uparrow}_{dp1\sigma} = -1.2154 \\
&V^{(3)\uparrow}_{dp1\pi} = 0.5003 \\
\end{split}
\end{equation}
and
\begin{equation}
\begin{split}
&V^{(1)\uparrow}_{dp2\sigma} = 0.0089 \\
&V^{(1)\uparrow}_{dp2\pi} = -0.0677 \\
&V^{(2)\uparrow}_{dp2\sigma} = -0.0141 \\
&V^{(2)\uparrow}_{dp2\pi} = -0.0424 \\
&V^{(3)\uparrow}_{dp2\sigma} = -0.0170 \\
&V^{(3)\uparrow}_{dp2\pi} = -0.0147 \\
\end{split}
\end{equation}
For spin-down,
\begin{equation}
\begin{split}
&V^{(1)\downarrow}_{dp1\sigma} = -4.7869 \\
&V^{(1)\downarrow}_{dp1\pi} = 0.4282 \\
&V^{(2)\downarrow}_{dp1\sigma} = -0.7743 \\
&V^{(2)\downarrow}_{dp1\pi} = 0.4665 \\
&V^{(3)\downarrow}_{dp1\sigma} = -1.1528 \\
&V^{(3)\downarrow}_{dp1\pi} = 0.3485 \\
\end{split}
\end{equation}
and
\begin{equation}
\begin{split}
&V^{(1)\downarrow}_{dp2\sigma} = 0.0226 \\
&V^{(1)\downarrow}_{dp2\pi} = -0.0295 \\
&V^{(2)\downarrow}_{dp2\sigma} = 0.0097 \\
&V^{(2)\downarrow}_{dp2\pi} = -0.0202 \\
&V^{(3)\downarrow}_{dp2\sigma} = -0.0036 \\
&V^{(3)\downarrow}_{dp2\pi} = -0.0134 \\
\end{split}
\end{equation}

Similarly, $d$-$d$ hoppings are constructed from blockwise parameters for blocks 1-4, but they do not need to be rotated. 
Because the Cr atoms are in the same plane, directional cosine $n$ vanishes, \textit{i.e.}, $n=0$. It causes ambiguities in the determination of the Slater-Koster parameters.
In block 1, $V^{(1)}_{dd\pi}$ does not appear and only $V^{(1)}_{dd\sigma}+3V^{(1)}_{dd\delta}$ is uniquely determined, not individually. To resolve the ambiguity, we set $V^{(1)}_{dd\pi} = V^{(1)}_{dd\delta} =0$. This approach is justified when we use the Slater-Koster parameters as a fitting rules obeying the symmetry and do not impart the physical meaning to each parameter.
In block 2, similarly, only $-V^{(2)}_{dd\sigma}+V^{(2)}_{dd\delta}$ is uniquely determined. We set $V^{(2)}_{dd\pi} = V^{(2)}_{dd\delta} =0$.
In block 3, $3V^{(3)}_{dd\sigma}+V^{(3)}_{dd\delta}$ and $V^{(3)}_{dd\pi}$ are uniquely determined. $V^{(3)}_{dd\delta}=0$ is enough.
In block 4, $V^{(4)}_{dd\sigma}$ does not appear, so $V^{(4)}_{dd\sigma}=0$.
From these settings, we obtain the following parameters in eV.
For spin-up,
\begin{equation}
\begin{split}
&V^{(1)\uparrow}_{dd\sigma} = -0.2283 \\
&V^{(1)\uparrow}_{dd\pi}    = 0 \\
&V^{(1)\uparrow}_{dd\delta} = 0 \\
&V^{(2)\uparrow}_{dd\sigma} = -0.2977 \\
&V^{(2)\uparrow}_{dd\pi}    = 0 \\
&V^{(2)\uparrow}_{dd\delta} = 0 \\
&V^{(3)\uparrow}_{dd\sigma} = -0.1407 \\
&V^{(3)\uparrow}_{dd\pi}    = 0.0849 \\
&V^{(3)\uparrow}_{dd\delta} = 0 \\
&V^{(4)\uparrow}_{dd\sigma} = 0 \\
&V^{(4)\uparrow}_{dd\pi}    = -0.0872 \\
&V^{(4)\uparrow}_{dd\delta} = -0.0054 \\
\end{split}
\end{equation}
For spin-down,
\begin{equation}
\begin{split}
&V^{(1)\downarrow}_{dd\sigma} = -0.0642 \\
&V^{(1)\downarrow}_{dd\pi}    = 0 \\
&V^{(1)\downarrow}_{dd\delta} = 0 \\
&V^{(2)\downarrow}_{dd\sigma} = -0.0779 \\
&V^{(2)\downarrow}_{dd\pi}    = 0 \\
&V^{(2)\downarrow}_{dd\delta} = 0 \\
&V^{(3)\downarrow}_{dd\sigma} = -0.0185 \\
&V^{(3)\downarrow}_{dd\pi}    = 0.0041 \\
&V^{(3)\downarrow}_{dd\delta} = 0 \\
&V^{(4)\downarrow}_{dd\sigma} = 0 \\
&V^{(4)\downarrow}_{dd\pi}    = -0.0790 \\
&V^{(4)\downarrow}_{dd\delta} = -0.0008 \\
\end{split}
\end{equation}

In addition, the fact $n=0$ makes the elements in the blocks 5 and 6 of $d$-$d$ matrix vanish.
However, the absence of $\mathcal{M}_z$ symmetry additionally allows non-zero elements for blocks 5 and 6, which can not be parametrized by the standard Slater-Koster parametrization. However, the remaining $D_3$ symmetry allows an alternative parametrization.
Let us consider two Cr atoms, one is at the origin, and the other is at $-d\hat{\mathbf{y}}$ where $d$ is Cr-Cr distance.
In our settings, one of the 2-fold rotation axes coincides with the $y$-axis.
The corresponding hopping term can be expressed as $\mel{d_{\alpha}(\mathbf{r})}{\hat{H}}{d_{\beta}(\mathbf{r}+d\hat{\mathbf{y}})} = P^{C_{2y}}_{\alpha}P^{C_{2y}}_{\beta}\mel{d_{\alpha}(\mathbf{r})}{\hat{H}}{d_{\beta}(\mathbf{r}+d\hat{\mathbf{y}})}$ where $P^{C_{2y}}_{\alpha}=\pm 1$ is parity of the $d_{\alpha}$ orbital under the 2-fold rotation $C_{2y}$.
Therefore, $P^{C_2}_{\alpha}P^{C_2}_{\beta}=1$ is the condition that the hopping does not vanish.
Let us define the non-vanishing terms as $\mel{d_{z^2}(\mathbf{r})}{\hat{H}}{d_{zx}(\mathbf{r}+d\hat{\mathbf{y}})}=V_1$,
$\mel{d_{xy}(\mathbf{r})}{\hat{H}}{d_{yz}(\mathbf{r}+d\hat{\mathbf{y}})}=V_2$, and
$\mel{d_{x^2-y^2}(\mathbf{r})}{\hat{H}}{d_{zx}(\mathbf{r}+d\hat{\mathbf{y}})}=V_3$.
Next, consider the hopping to the Cr in another site, which can be represented by the counter-clockwise relative angle $\phi$ with respect to $-d\hat{\mathbf{y}}$. The orbitals in this site are equivalent to the orbitals located at $-d\hat{\mathbf{y}}$ but rotated by $-\phi$. The corresponding rotation rules can be expressed as follows.
\begin{equation}
\begin{split}
x' &=  \cos\phi x -\sin\phi y = \cos(\tfrac{\pi}{2}+\theta)x-\sin(\tfrac{\pi}{2}+\theta)y \\
&= -\sin\theta x -\cos\theta y = -mx-ly \\
y' &= \sin\phi x +\cos\phi y = \sin(\tfrac{\pi}{2}+\theta)x+\cos(\tfrac{\pi}{2}+\theta)y \\
&= \cos\theta x -\sin\theta y = lx-my
\end{split}
\end{equation}
where $\theta$ is an angle measured from the $x$-axis.
By applying these rules to Eq.~(\ref{eq:dorbitals}), one can obtain the following parametrization.
\begin{equation}
\mel{d_{i,z^2}}{\hat{H}}{d_{j,zx}}= - \chi^i mV_{1}
\end{equation}
\begin{equation}
\mel{d_{i,z^2}}{\hat{H}}{d_{j,yz}}= \chi^i lV_{1}
\end{equation}
\begin{equation}
\mel{d_{i,x^2-y^2}}{\hat{H}}{d_{j,zx}}= \chi^i(-2l^2mV_2 + (l^2-m^2)m V_3)
\end{equation}
\begin{equation}
\mel{d_{i,x^2-y^2}}{\hat{H}}{d_{j,yz}}= \chi^i(-2lm^2V_2 - (l^2-m^2)l V_3)
\end{equation}
\begin{equation}
\mel{d_{i,xy}}{\hat{H}}{d_{j,zx}}= \chi^i((l^2-m^2)l V_2 + 2lm^2V_3)
\end{equation}
\begin{equation}
\mel{d_{i,xy}}{\hat{H}}{d_{j,zx}}= \chi^i((l^2-m^2)m V_2 - 2l^2mV_3)
\end{equation}
where $\chi^i=\pm 1$ represents the OC of Cr-$i$, $+1$ for R and $-1$ for L. OC dependency in this parametrization comes from the fact that R-OC and L-OC are related by $\mathcal{M}_z$ whose breaking is the origin of these terms.
For spin-up, obtained parameters are
\begin{equation}
\begin{split}
&V^{\uparrow}_{1} = -0.0567 \\
&V^{\uparrow}_{2} = -0.0717 \\
&V^{\uparrow}_{3} = -0.0484 \\
\end{split}
\end{equation}
and for spin-down,
\begin{equation}
\begin{split}
&V^{\downarrow}_{1} = -0.0259 \\
&V^{\downarrow}_{2} = -0.0380 \\
&V^{\downarrow}_{3} = -0.0262 \\
\end{split}
\end{equation}

\begin{figure}[t!]
\centering
\includegraphics[width=0.5\textwidth]{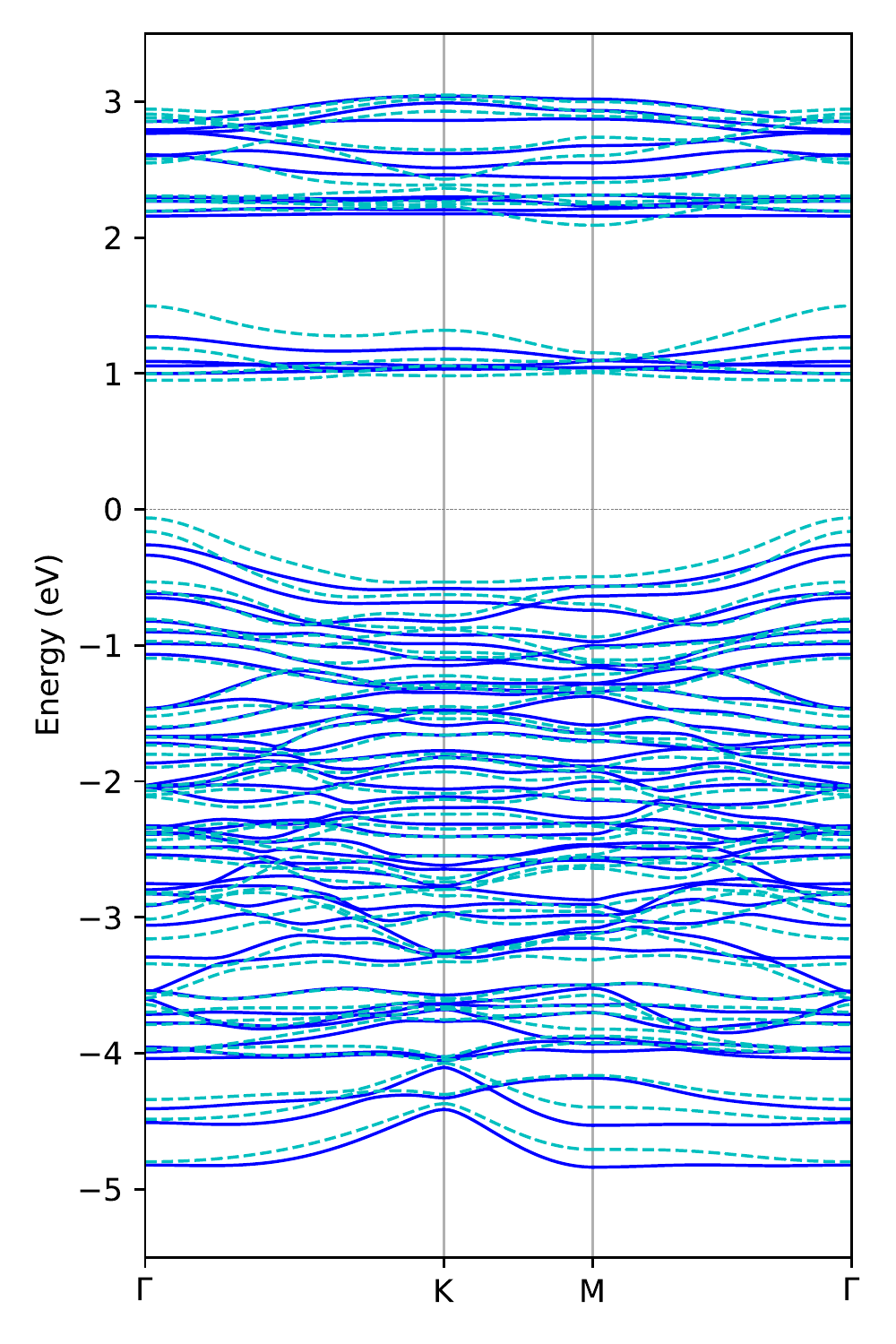}
\caption{
Band structure of monolayer. Solid lines are bands calculated by the TB model, and dashed lines are bands calculated by DFT.}
\label{fig:bands_mono}
\end{figure}

\begin{figure}[t!]
\centering
\includegraphics[width=0.5\textwidth]{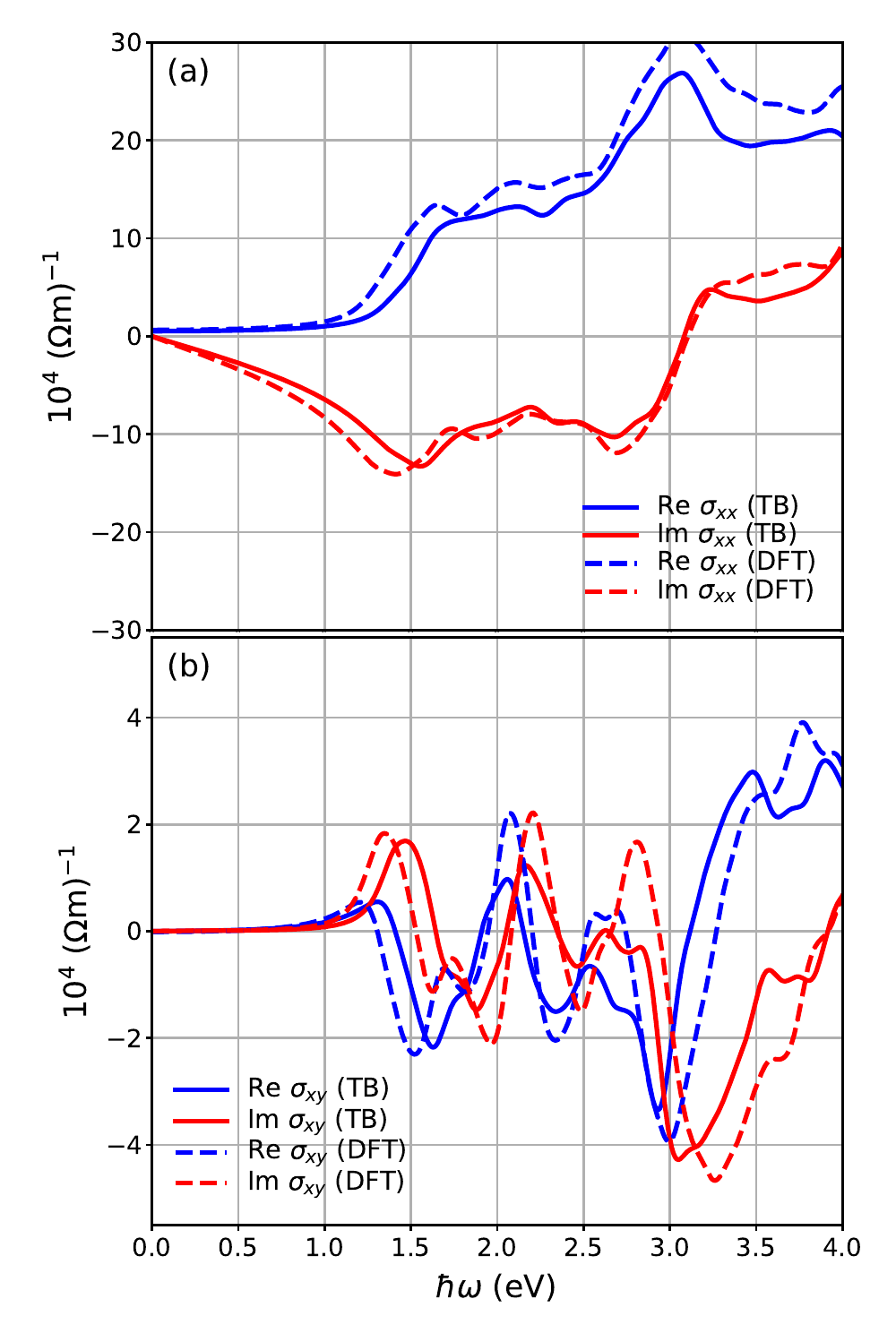}
\caption{
Conductivity tensor components calculated from TB and DFT (Wannier interpolation). (a) diagonal component and (b) off-diagonal component.}
\label{fig:sigma}
\end{figure}

\subsection{TB Results of Monolayer}
In Fig.~\ref{fig:bands_mono}, the band structure of the monolayer CrI$_3$ calculated from the TB model Hamiltonian is shown. For comparison, the band structure from DFT is shown together. TB bands show a good agreement with the DFT bands. The valence band top is at the $\Gamma$ point. The lowest conduction band is considerably flat.
It can be noted that the increased band gap in TB band leads to the positive shift of the first Kerr angle peak. From DFT, gaps at the $\Gamma$, $K$, and $M$ points are 1.011, 1.516, and 1.502 eV, respectively. On the other hand, those from the TB model are  1.260, 1.614, and 1.606 eV, respectively.

Fig.~\ref{fig:sigma} shows the conductivity tensor components $\sigma_{xx}$ and $\sigma_{xy}$ calculated by TB model and Wannier interpolation.
The $\sigma_{xx}$ exhibits a good agreement in both the real and imaginary parts in the given range.
Despite the shift of the peak position due to the increased gap, the off-diagonal component $\sigma_{xy}$ by the TB model also well reproduces the Wannier results in its behavior and value. These are qualitatively consistent with the literature [S5, S6].
In addition to the MOKE spectrum in the main text, these results justify our TB model when we investigate the low energy properties.

\begin{figure}[t!]
\centering
\includegraphics[width=0.5\textwidth]{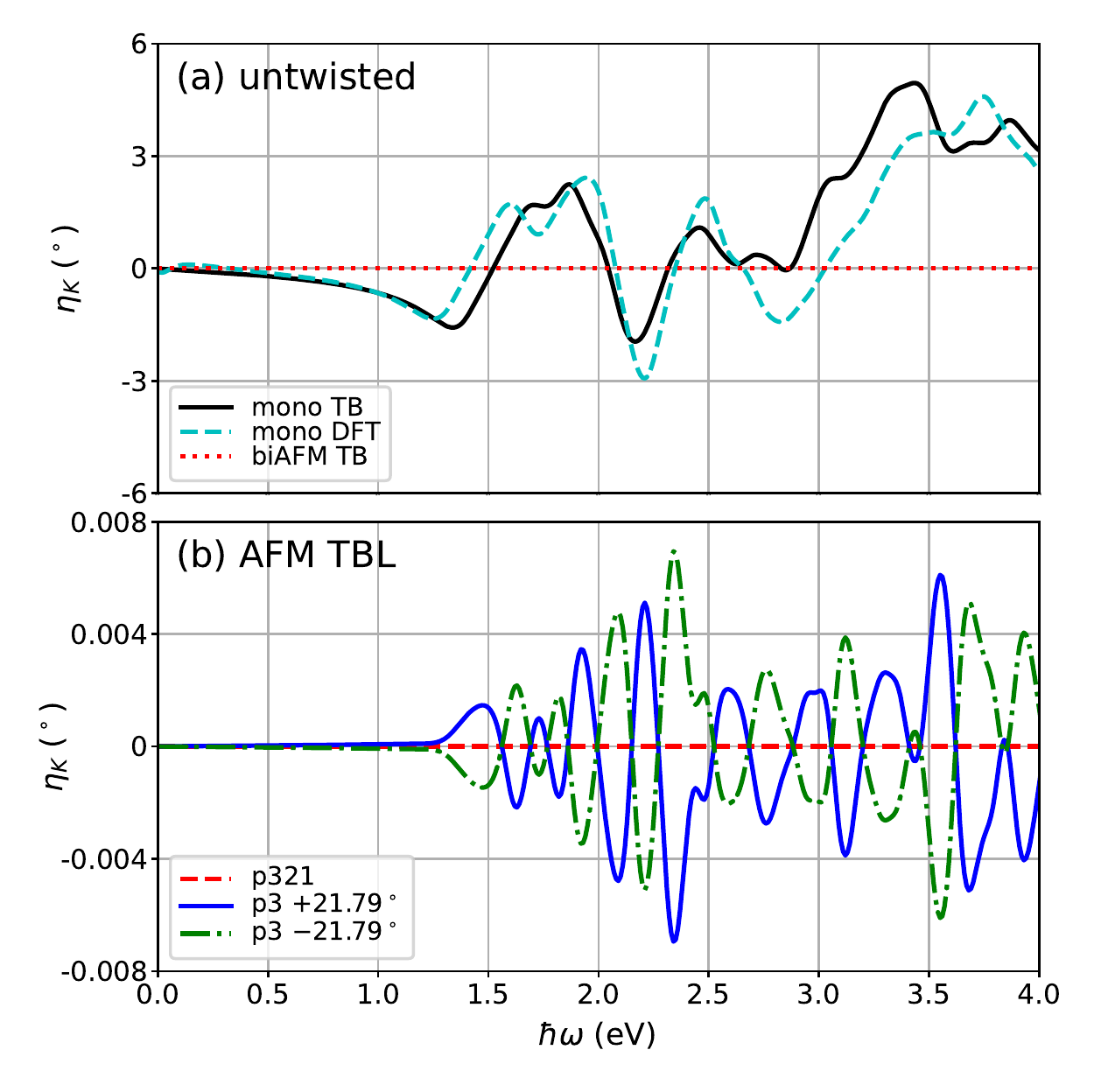}
\caption{
Kerr ellipticity in degrees with respect to the photon energy $\hbar\omega$ of
(a) FM monolayer and untwisted AFM bilayer, and (b) AFM TBLs CrI$_3$.}
\label{fig:MOKE_ellipticity}
\end{figure}

\begin{figure}[t!]
\centering
\includegraphics[width=0.5\textwidth]{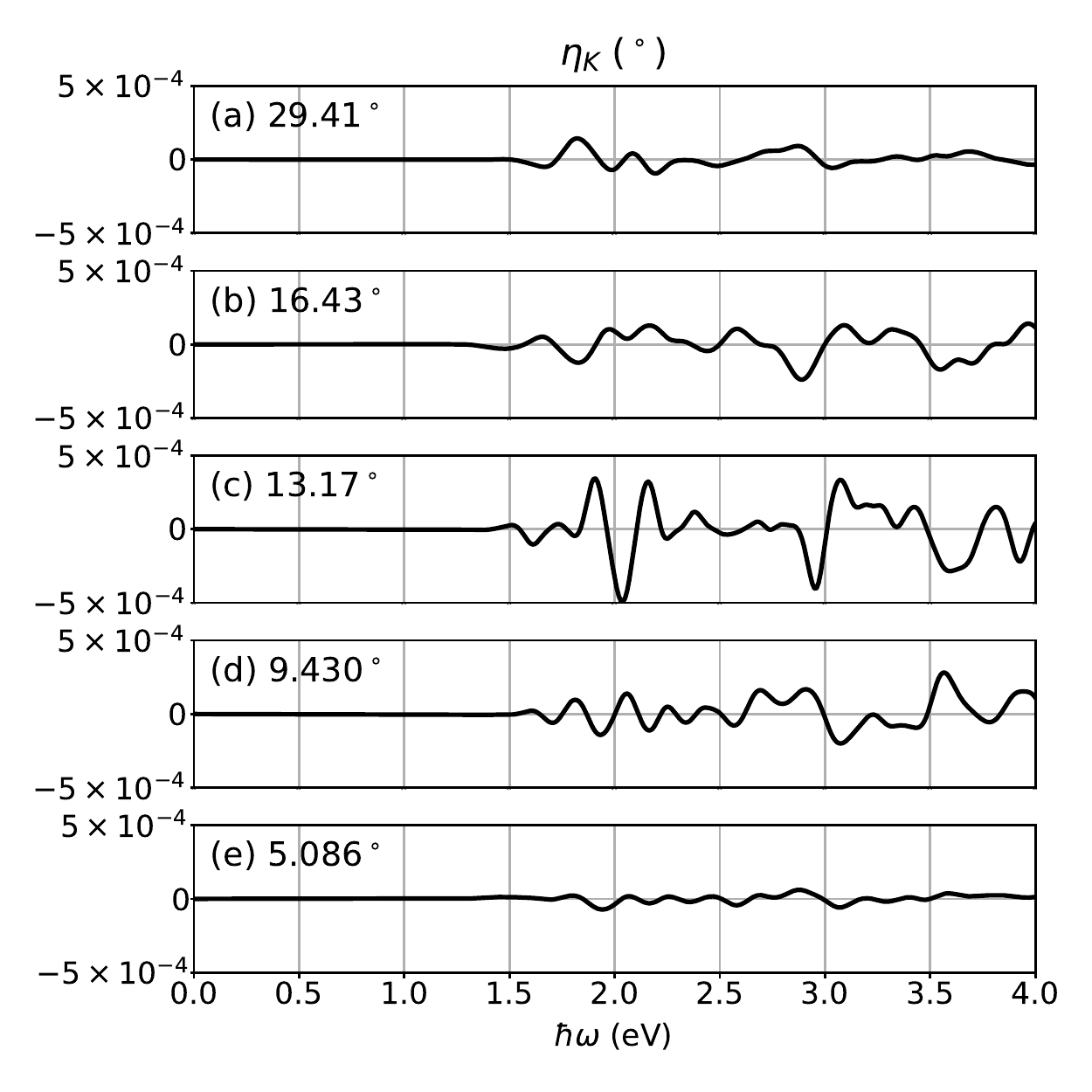}
\caption{
Kerr ellipticity spectra of the systems with various twist angles. (a) 29.41$^\circ$, (b) 16.43$^\circ$, (c) 13.17$^\circ$, (d) 9.430$^\circ$, and (e) 5.086$^\circ$.}
\label{fig:MOKE_angle_ellipticity}
\end{figure}

\subsection{Opposite Octahedral Chirality}
In order to construct the Hamiltonian of the $p3$ TBL structure, the Hamiltonian of the monolayer with the inverted OCs is needed, \textit{i.e.}, Cr1 is R-OC and Cr2 is L-OC. This structure is obtained by $\mathcal{M}_z$ mirror operation to the original structure in Table~\ref{tab:fractional_coord}.
Also, in this case, local coordinates for $p$-orbitals are rotated by $180^\circ$ (as a result, rotation angles are $30^\circ$ for I3 and I8, $150^\circ$ for I4 and I6, and $-90^\circ$ for I5 and I7), so that the $p$-orbitals seen by the Cr with a specific OC are the same in both types of monolayers. In the new local reference frame, $h_{ppN}$s obtained previously can be used as they are because the $\mathcal{M}_z$ changes the sign of $z$ and the $180^\circ$ rotation changes the sign of $x$ and $y$, \textit{i.e.}, twice of sign changes for every combination of the $p$-$p$ hopping. 
Because the local environment around Cr is changed by $\mathcal{M}_z$, $\delta_d$ in the intra-atomic $d$-orbital Hamiltonian should change its sign. However, the sign change in the $d$-$d$ hopping by $\mathcal{M}_z$ breaking effect is already reflected by $\chi^i$ explicitly.

\subsection{Interlayer Coupling}
Interlayer coupling is described by the $p$-$p$ hopping between the bottom layer's upper I and the top layer's lower I. It is determined by Slater-Koster parametrization with the rotation rules described in a previous subsection considering the local coordinates. Interaction strength follows the exponentially decay rule [S7, S8],
\begin{equation}
V^{\text{IL}}_{pp\sigma/pp\pi}(r) = V^{\text{IL}0}_{pp\sigma/pp\pi}\exp(-\frac{r-r_{\text{ref}}}{r_{\text{scale},pp\sigma/pp\pi}})
\end{equation}
where $r$ is inter-atomic distance.
Parameters are obtained by fitting to the Wannier Hamiltonian obtained from the DFT calculations for the bilayers of AB stacked, and those with the top layers shifted by $\pm0.15\mathbf{a}$. In fact, $r_{\text{ref}}$ is a redundant parameter. However, this is introduced to make $V^{\text{IL}0}_{pp\sigma/pp\pi}$ have a reasonable value and chosen as the interplane distance between two iodine planes.
\begin{equation}
\begin{split}
&V^{\text{IL}0}_{pp\sigma} = 0.4112\ \text{eV}\\
&V^{\text{IL}0}_{pp\pi} = -0.0603\ \text{eV}\\
&r_{\text{ref}} = 4.1284\ \text{\AA}\\
&r_{\text{scale},pp\sigma} = 0.6464\ \text{\AA} \\
&r_{\text{scale},pp\pi} = 0.4446\ \text{\AA} \\
\end{split}
\end{equation}
In practice, only the interlayer couplings that the interaction distance is shorter than $r_{\text{cut}} = a$ are included in the calculation.

\section{Kerr Ellipticity}
In this supplemental section, Kerr ellipticity $\eta_{\text{K}}$ is shown. Kerr ellipticity measures how much the reflected light becomes circularly polarized.
Fig~\ref{fig:MOKE_ellipticity} (a) shows the $\eta_{\text{K}}$ of the monolayer calculated by the TB and Wannier function, respectively, and of the AFM bilayer by the TB. Kerr ellipticity is also well reproduced by TB.
Fig~\ref{fig:MOKE_ellipticity} (b) shows the $\eta_{\text{K}}$ of TBL systems confirming the predictions by the symmetry.
Fig~\ref{fig:MOKE_angle_ellipticity} shows the twist angle dependence of the Kerr ellipticity.

\section{Spin Textures}
\begin{figure*}[t!]
\centering
\includegraphics[width=0.8\textwidth]{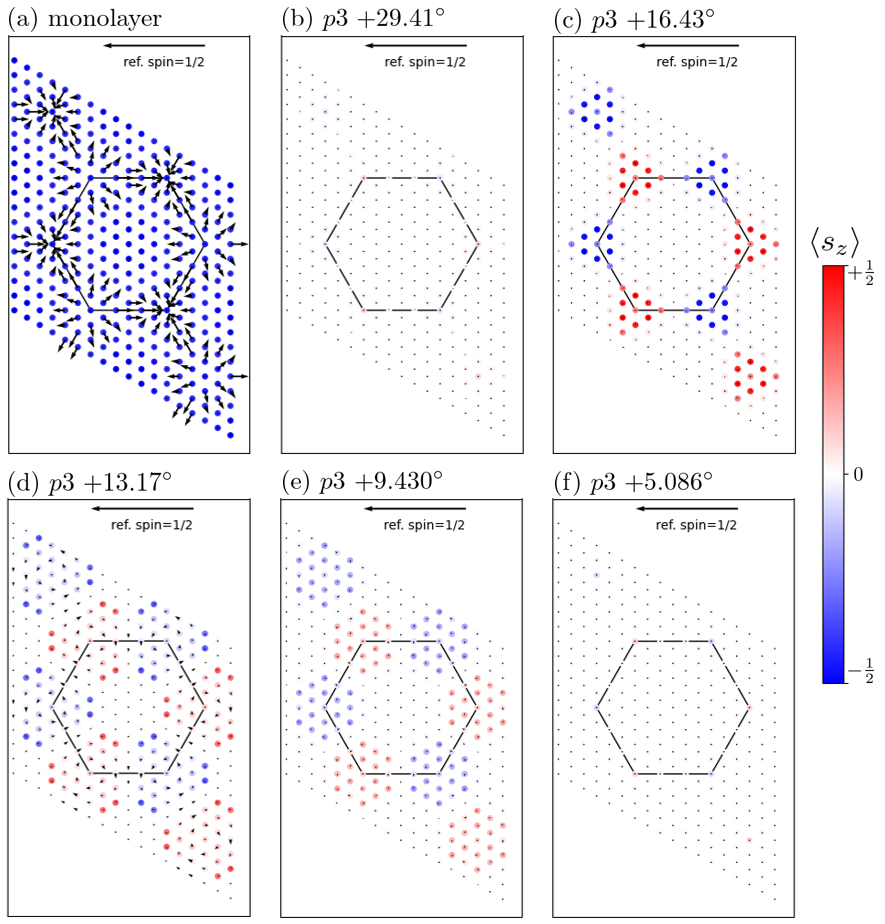}
\caption{
Spin textures of the highest valence band of the FM monolayer and the AFM twisted bilayers in the $k$-space. (a) monolayer. $p3$ structure of (b) $+29.41^\circ$, (c) $+16.43^\circ$, (d) $+13.17^\circ$, (e) $+9.430^\circ$, and (f) $+5.086^\circ$ twist.}
\label{fig:spintexture_angle}
\end{figure*}

Spin textures of the highest valence bands are shown in Fig.~\ref{fig:spintexture_angle} for the FM monolayer and the AFM $p3$ systems of the twist angles in Table~\textrm{III} of the main text except for $21.79^\circ$.
In the monolayer [Fig.~\ref{fig:spintexture_angle} (a)], spins are almost polarized to $-z$ direction. However, small canting in the radial direction is found around the $K$ points.
In the $p3$ systems, only the $13.17^\circ$ case [Fig.~\ref{fig:spintexture_angle} (d)] exhibits a tiny in-plane helical spin texture, while the other cases show almost no in-plane spin textures.
Moreover, the out-of-plane spin textures are almost vanishing in the $29.41^\circ$ and $5.086^\circ$ cases [Fig.~\ref{fig:spintexture_angle} (b) and (f)].

\begin{figure*}[t!]
\centering
\includegraphics[width=0.8\textwidth]{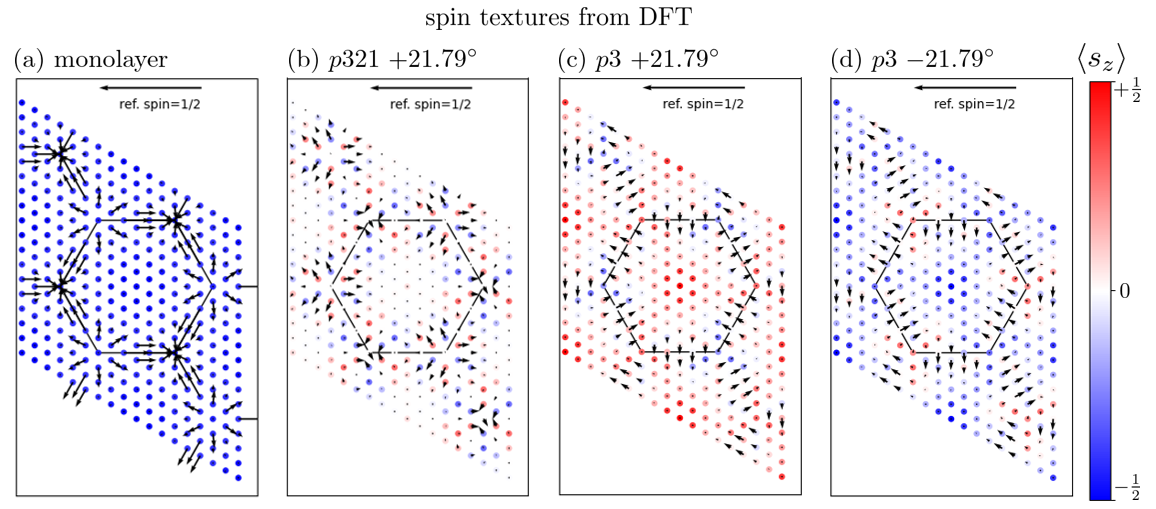}
\caption{
Spin textures calculated by DFT. (a) FM monolayer, (b) $p321$ structure $+21.79^\circ$, and $p3$ structure of (c,d) $\pm21.79^\circ$.}
\label{fig:spintexture_DFT}
\end{figure*}

Finally, we cross-checked the spin texture by DFT. Fig.~\ref{fig:spintexture_DFT} shows the spin textures of the monolayer and the TBL systems with $\pm21.79^\circ$ twist angle calculated from DFT, which are the counterparts of Fig.~\ref{fig:spintexture_angle} (a) and Fig. 8 of the main text, respectively.
In each case, the characteristics of the in-plane spin textures are in good agreement with the TB model results. Out-of-plane spin textures of the monolayer and $p321$ TBL system also show a good agreement. However, in $p3$ TBL systems, the signs of the out-of-plane component around the Brillouin zone center ($\Gamma$) are opposite to the TB model spin texture, whereas those around the $K$ points are consistent with the TB model.
Note that the deviation between the DFT and TB bands is more significant around $\Gamma$ point than around the $K$ points [See Fig.~\ref{fig:bands_mono}]. This could explain the sign discrepancy occurring around $\Gamma$ point for the spin texture.
Despite this mismatch, the switching rule between the enantiomers of the $p3$ systems [$\mathbf{s}^{+\theta}_{\mathbf{k}} = (s_x,s_y,s_z)$ and $\mathbf{s}^{-\theta}_{-\mathbf{k}} = (s_x,s_y,-s_z)$] is also valid in the DFT results.

\hfill

\textbf{References for SM}

\hfill

[S1] N. Marzari and D. Vanderbilt, Phys. Rev. B 56, 12847 (1997).

[S2] T. Yusufaly, D. Vanderbilt, and S. Coh, (2018), Tight-Binding Formalism in the Context of the PythTB Package, https://www.physics.rutgers.edu/pythtb/formalism.html.

[S3] M. Graf and P. Vogl, Phys. Rev. B 51, 4940 (1995).

[S4] J. C. Slater and G. F. Koster, Phys. Rev. 94, 1498 (1954).

[S5] V. Kumar Gudelli and G.-Y. Guo, New J. Phys. 21, 053012 (2019).

[S6] A. A. Pervishko, D. Yudin, V. Kumar Gudelli, A. Delin, O. Eriksson, and G.-Y. Guo, Opt. Express 28, 29155 (2020).

[S7] G. Trambly de Laissardi\'ere, D. Mayou, and L. Magaud, Nano Lett. 10, 804 (2010).

[S8] S. Fang, R. Kuate Defo, S. N. Shirodkar, S. Lieu, G. A. Tritsaris, and E. Kaxiras, Phys. Rev. B 92, 205108 (2015).

\end{document}